\documentclass[letter,11pt]{article}
\usepackage{jheppub}
\usepackage{tikz}
\usetikzlibrary{decorations.pathmorphing}
\usetikzlibrary{decorations.markings}
\usepackage{xspace}
\usepackage{amsmath}
\usepackage{amssymb}
\usepackage{mathtools}
\usepackage[utf8]{inputenc}
\usepackage{physics}
\usepackage{slashed}
\usepackage{hyperref}
\usepackage[capitalize]{cleveref}

\newcommand{\g}{\gamma}
\def\2g{{\gamma\gamma}}
\def\j2g{{\gamma\gamma+{\rm jet}}}
\def\3g{{\gamma\gamma\gamma}}
\def\pt2g{{p_T(\g\g)}}
\def\mgg{{m(\g\g)}}

\title{NNLO QCD corrections to diphoton production with an additional jet at the LHC}

\author[a]{Herschel A. Chawdhry,}
\author[b]{Micha\l{}  Czakon,}
\author[c]{Alexander Mitov,}
\author[c]{Rene Poncelet}

\affiliation[a]{Rudolf Peierls Centre for Theoretical Physics, Clarendon Laboratory, University of Oxford, Oxford OX1 3PU, United Kingdom}
\affiliation[b]{Institut f\"ur Theoretische Teilchenphysik und Kosmologie, RWTH Aachen University, D-52056 Aachen, Germany}
\affiliation[c]{Cavendish Laboratory, University of Cambridge, Cambridge CB3 0HE, United Kingdom}

\note{Preprint: Cavendish-HEP-21/07, OUTP-21-13P, P3H-21-032}

\abstract{We calculate the NNLO QCD corrections to diphoton production with an additional jet at the LHC. Our calculation represents the first NNLO-accurate prediction for the transverse momentum distribution of the diphoton system. The improvement in the accuracy of the theoretical prediction is significant, by a factor of up to four relative to NLO QCD. Our calculation is exact except for the finite remainder of the two-loop amplitude which is included at leading color. The numerical impact of this approximated contribution is small. The results of this work are expected to further our understanding of the Higgs boson sector and of the behavior of higher-order corrections to LHC processes.}

\begin{document} 
\maketitle
\flushbottom

\section{Introduction}\label{sec:intro}

The production of a pair of photons at the LHC is of special interest. On the one hand this process represents the main background to the cleanest Higgs boson decay channel $h\to\2g$. On the other hand it is a process where our ability to accurately predict LHC cross sections by including higher-order QCD corrections can reliably be tested. 

Inclusive diphoton production $pp\to\2g+X$ has been studied extensively at NNLO in QCD \cite{Catani:2011qz,Campbell:2016yrh,Catani:2018krb,Gehrmann:2020oec,Alioli:2020qrd}. Prior work also includes NLO QCD corrections \cite{DelDuca:2003uz} or electroweak effects \cite{Chiesa:2017gqx}, as well as photon isolation \cite{Gehrmann:2013aga}. Beyond NLO, resummation effects have been included at least to NNLL \cite{Cieri:2015rqa,Coradeschi:2017zzw,Becher:2020ugp,Ebert:2016gcn}. Interference between $\j2g$ and $h\to\2g+j$ have also been investigated \cite{Cieri:2017kpq}. A detailed analysis of the limitations of existing results can be found in ref.~\cite{Amoroso:2020lgh}.

A distinguishing feature of this process is the presence of very large higher-order QCD corrections which has raised some questions about the reliability of such higher-order predictions for this process. With time our understanding of the behavior of higher-order corrections in this case has developed significantly. It is presently believed \cite{Catani:2018krb} that starting with N$^3$LO, perturbative corrections will be much more mild and consistent with perturbative convergence. A similar conclusion has also been reached recently for the process $pp\to\3g$ \cite{Chawdhry:2019bji,Kallweit:2020gcp}. Clearly it is very desirable to have a full N$^3$LO accurate calculation of inclusive diphoton production where these ideas can be tested and hopefully validated. The present work, together with the recently computed 3-loop amplitudes for diphoton production at the LHC \cite{Caola:2020dfu}, represents a significant step in this direction.

The transverse momentum of the photon pair, $\pt2g$, plays a special role in inclusive diphoton production $pp\to\2g+X$. As is well known, due to the fact that at leading order $\pt2g=0$, an NNLO-accurate calculation of inclusive diphoton production is only NLO-accurate for the $\pt2g$ distribution. To achieve NNLO accuracy for $\pt2g$ at nonzero $\pt2g$ one needs to compute the NNLO QCD corrections for the process $pp\to\j2g+X$. The present work presents the first calculation of $pp\to\j2g+X$ in NNLO QCD and makes public NNLO QCD predictions for a number of diphoton observables at nonzero $\pt2g$. 

The main reason the $p_T$ distribution of the diphoton system is of special interest is that it represents the main background for Higgs production at high $p_T$. High-$p_T$ Higgs production is relevant for Dark Matter searches \cite{Sirunyan:2018fpy,Aad:2021qks} and for disentangling the nature of the Higgs boson's local vertex \cite{Lindert:2018iug} which is not possible at low $p_T$ where the effective $ggh$ vertex describes Higgs production well. The $\pt2g$ distribution, together with the angular distribution of the two photons in the Collins-Soper frame \cite{Collins:1977iv}, represents a strong discriminator for the spin of a possible resonance decaying to two photons \cite{Aad:2015mxa}. For further details about high-$p_T$ Higgs production we refer the reader to the recent review \cite{Becker:2020rjp}.

This work is organized as follows: in sec.~\ref{sec:setup} we briefly describe our calculation while in sec.~\ref{sec:pheno} we present our predictions for a number of differential distributions. Our conclusions are given in sec.~\ref{sec:conclusions}.

\section{Setup of the Calculation}\label{sec:setup}

The calculation is performed in the {\tt STRIPPER} approach \cite{Czakon:2010td,Czakon:2011ve,Czakon:2014oma}. The approach has already been applied in the calculation of NNLO QCD corrections to top-pair \cite{Czakon:2014xsa,Czakon:2015owf,Czakon:2016ckf,Czakon:2016dgf,Behring:2019iiv}, inclusive jet \cite{Czakon:2019tmo}, three-photon \cite{Chawdhry:2019bji}, $W+c ~{\rm jet}$ \cite{Czakon:2020coa}, identified $B$-hadron \cite{Czakon:2021ohs} and polarized $W$-pair \cite{Poncelet:2021jmj} production at the LHC. A detailed description of the technical aspects of our implementation has been given in ref.~\cite{Czakon:2019tmo}.

All tree-level diagrams are computed with the {\tt avhlib} library \cite{avhlib,Bury:2015dla}. The contributing one-loop amplitudes, including the loop-induced contribution, are obtained from the library {\tt OpenLoops} \cite{Cascioli:2011va,Buccioni:2019sur}.

The relevant two-loop contributions $q\bar q \to g\2g$ and $qg \to q\2g$ are handled in the following way. We first separate the finite remainders ${\cal H}^{(2)}(\mu_R^2)$ of the two-loop amplitudes from their infrared poles. The latter can be predicted exactly and we have included them, including their finite contributions, without any approximation. 
In terms of the scale dependence of the two-loop finite remainder ${\cal H}^{(2)}$ (defined as in ref.~\cite{Czakon:2014oma}):
\begin{eqnarray}
{\cal H}^{(2)}(\mu_R^2) ={\cal H}^{(2)}(s_{12})+\sum_{i=1}^4 c_i \ln^i\left({\mu_R^2\over s_{12}}\right)\,,
\label{eq:amp-scale}
\end{eqnarray}
where $s_{12}$ is the squared partonic center-of-mass energy, we have included without any approximation all two-loop terms $c_i$ corresponding to $\ln^i(\mu_R^2)$ with $i \geq1$. The scale-independent part ${\cal H}^{(2)}(s_{12})$ is included in the leading color approximation as derived in ref.~\cite{Chawdhry:2021mkw} with the help of refs.~\cite{Chawdhry:2018awn,Chicherin:2020oor,Chawdhry:2020for} (an equivalent expression for the spin-averaged two-loop squared amplitude has also been derived in ref.~\cite{Agarwal:2021grm}). This is the only approximation made in this paper. Further details about the implementation of the leading color approximation of the two-loop finite remainder can be found in ref.~\cite{Chawdhry:2019bji}.

As a justification of the leading color approximation just described we have verified that the numerical contribution of the scale-independent part of the two-loop finite remainder ${\cal H}^{(2)}(s_{12})$ is small. For all distributions computed here it is about 1-2\% of the complete NNLO prediction in all bins. This makes it smaller than the Monte Carlo integration error in the differential distributions. More details can be found in sec.~\ref{sec:pheno}.

We have also included the so-called loop-induced (LI) contribution $gg\to g\2g$ which begins to contribute starting at NNLO. As we explain in sec.~\ref{sec:pheno} its effect is about five percent and is strongly dependent on the distribution.

The QED coupling is taken to be $\alpha=1/137$. The strong coupling constant and parton distributions are renormalized such that they evolve with $n_f=5$ active flavors. Diagrams with top-quark loops are included in all contributions except the two-loop finite remainder and the one-loop squared contribution. The current calculation uses the NNPDF3.1 pdf set \cite{Ball:2017nwa} of order that matches the order of the perturbative calculation. The value of the strong coupling constant is taken from the LHAPDF interface \cite{Buckley:2014ana}. The central values of the factorization and renormalization scales have been fixed to:
\begin{equation}
\mu_F^2=\mu_R^2 = \frac{1}{4}\left(m^2(\g\g)+ p_T(\g\g)^2\right)\,.
\label{eq:scale}
\end{equation}
This scale choice has been motivated by the scales used in refs.~\cite{DelDuca:2003uz,Gehrmann:2013aga}: we have replaced the $p_T$ of the jet with that of the photon pair since in our setup, see below, we have no explicit jet requirements. At LO the two scales are equivalent.

Scale uncertainty has been estimated with the help of a 7-point restricted independent variation by a factor of 2 of the scales $\mu_F$ and $\mu_R$. Since in this work we are primarily concerned with perturbative convergence and estimates of missing higher-order corrections, we have not included pdf error estimates. We expect that those are not dominant over the scale variation in the kinematic ranges considered here. We hope to include them in a future update of the present work. 

Our calculation has been performed for the LHC at 13 TeV and is subject to the following set of selection cuts:
\footnote{This set of cuts is based on typical selection requirements, see refs.~\cite{Aad:2015mxa,Sirunyan:2019iwo}.}
we require two photons satisfying the following criteria
    \begin{itemize}
     \item $p_T(\g_1) > 30$ GeV, $p_T(\g_2) > 18$ GeV and $|\eta(\g)| < 2.4\,,$
      \item Smooth photon isolation \cite{Frixione:1998jh} with $\Delta R_0 = 0.4$ and $E_T^{\rm max} =10~{\rm GeV}$ (see ref.~\cite{Chawdhry:2019bji} for details)\,,
      \item $m_{\g \g} \geq 90~ \text{GeV}\,,$
      \item $\Delta R (\g,\g) > 0.4\,,$
      \item $p_T(\g\g) > 20$ GeV (for lower values resummation effects become important \cite{Becher:2020ugp}) .
    \end{itemize}
No additional jet requirements are imposed. In particular, infrared safety is ensured by the $p_T(\g\g)$ cut specified above.

\section{Phenomenological results}\label{sec:pheno}

\begin{figure}[t]
\begin{centering}
\includegraphics[width = 0.48\textwidth,trim=0 1mm 0 0]{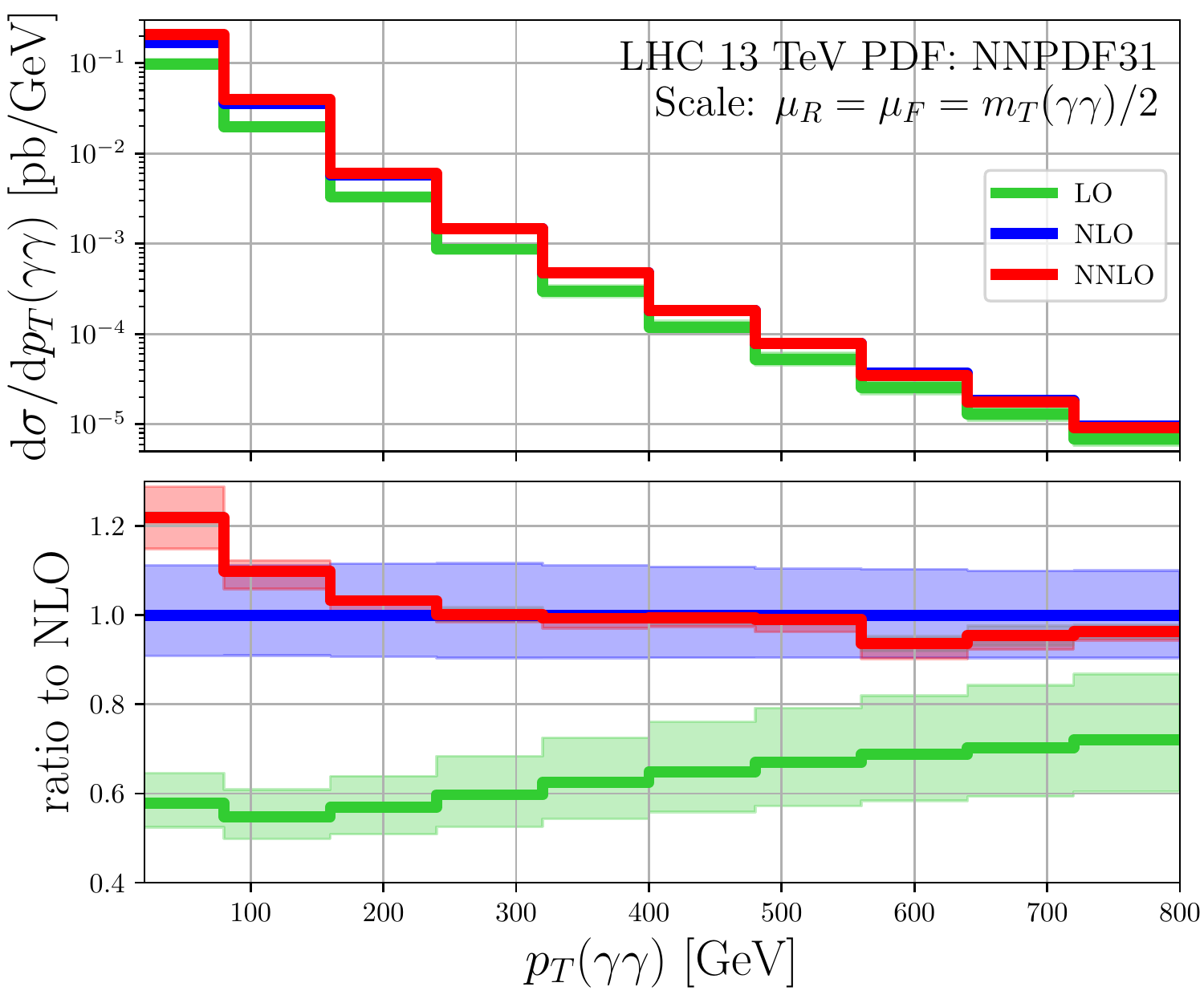}
\includegraphics[width = 0.48\textwidth,trim=0 1mm 0 0]{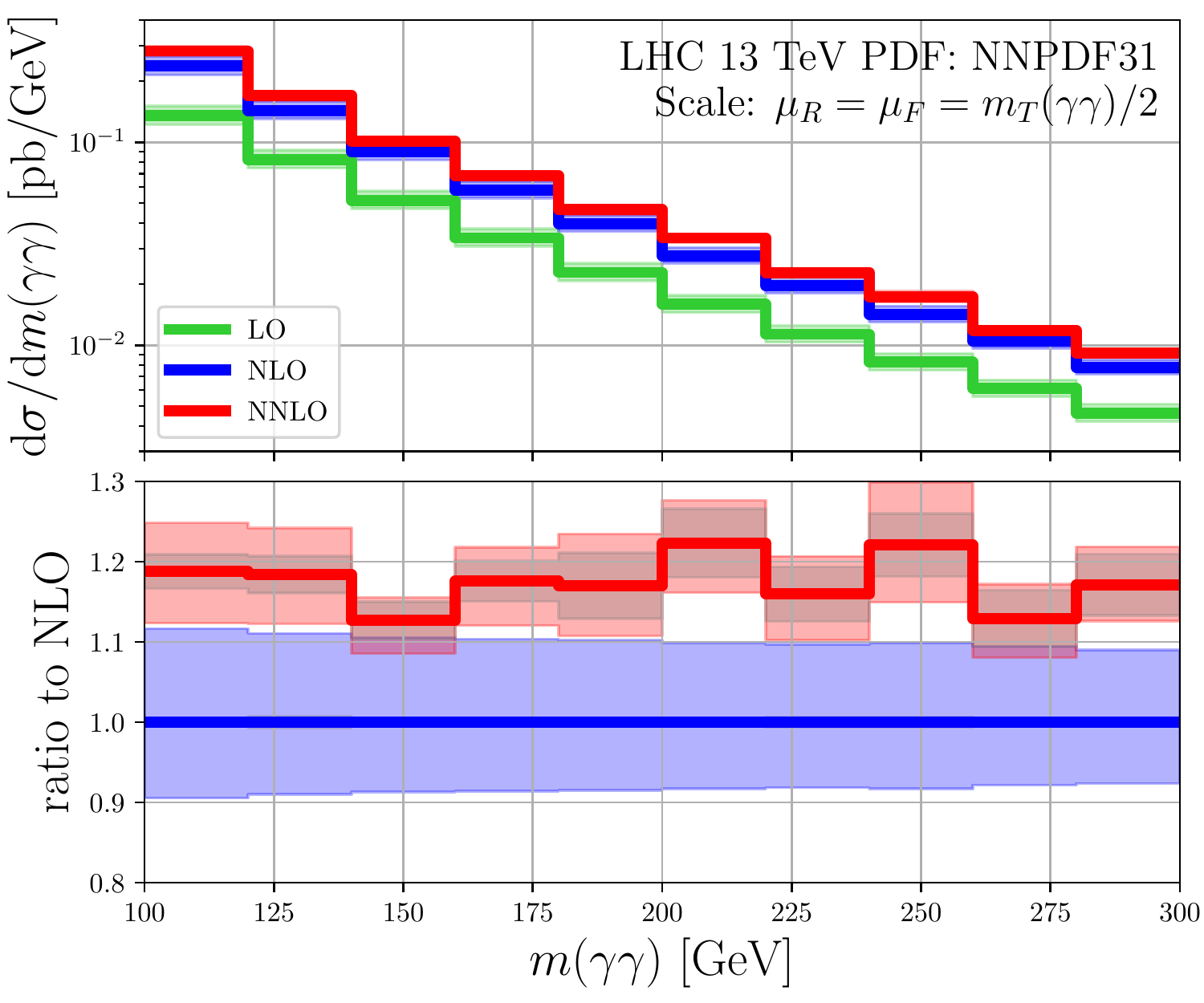}
\caption{Absolute $\pt2g$ (left) and $m(\g\g)$ (right) differential distributions. Shown are the predictions in LO (green), NLO (blue), NNLO (red) QCD. The colored bands around the central scales are from 7-point scale variation. The grey band shows the estimated Monte Carlo integration error in each bin. The lower panel shows the same distributions but relative to the NLO central scale prediction.}
\label{fig:PT-m}
\end{centering}
\end{figure}
\begin{figure}[t]
\begin{centering}
\includegraphics[width = 0.31\textwidth,trim=0 1mm 0 0]{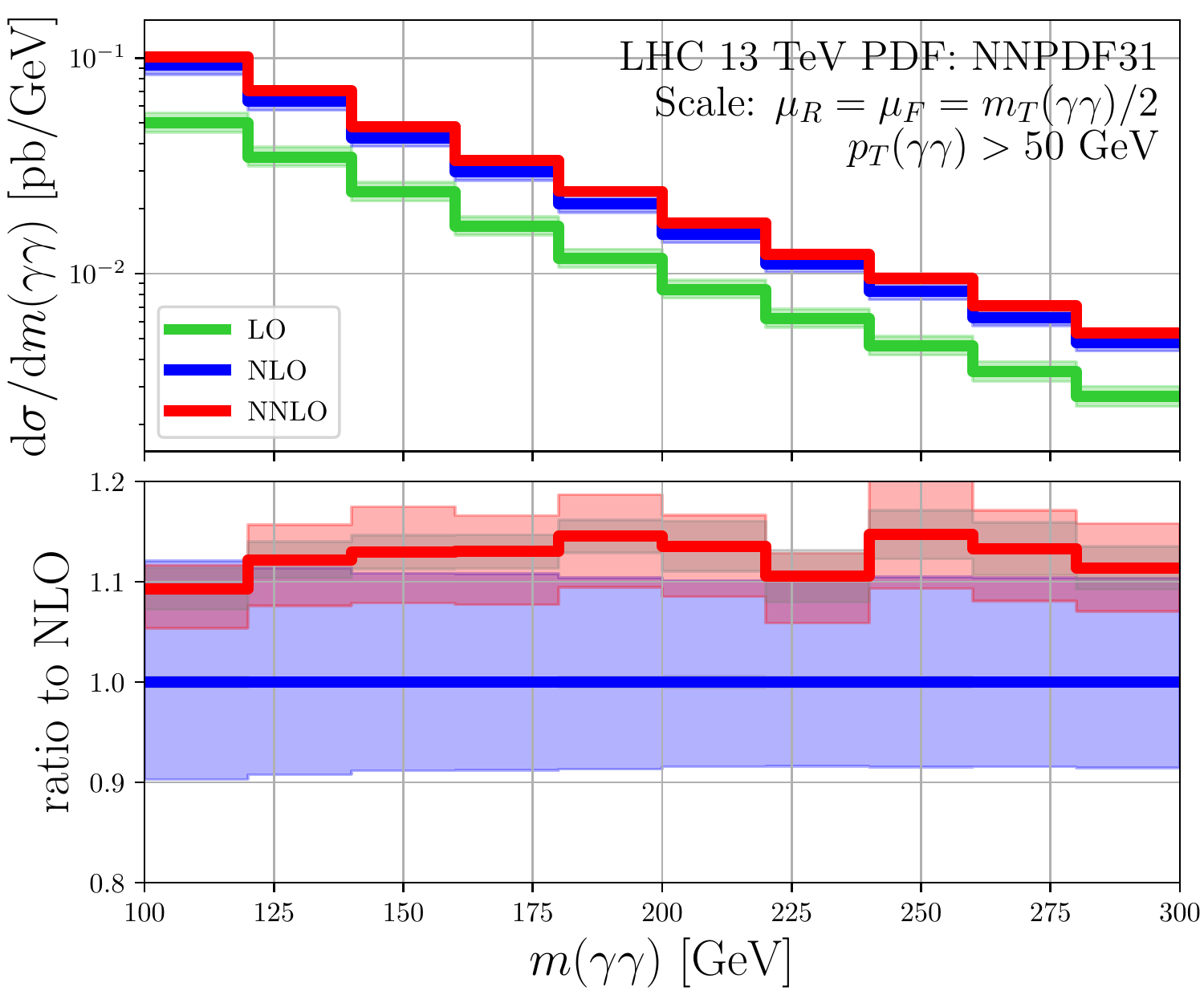}
\includegraphics[width = 0.31\textwidth,trim=0 1mm 0 0]{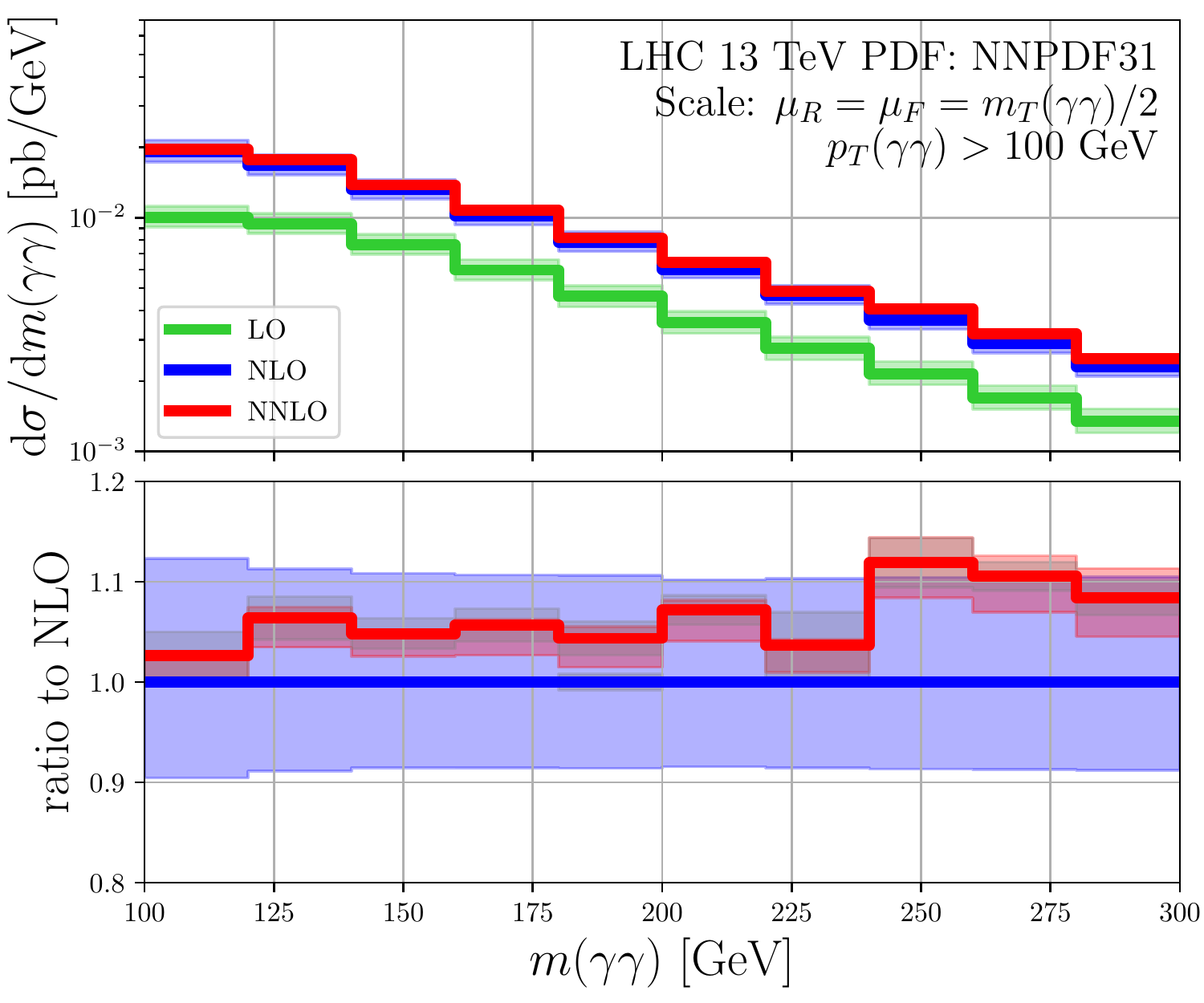}
\includegraphics[width = 0.31\textwidth,trim=0 1mm 0 0]{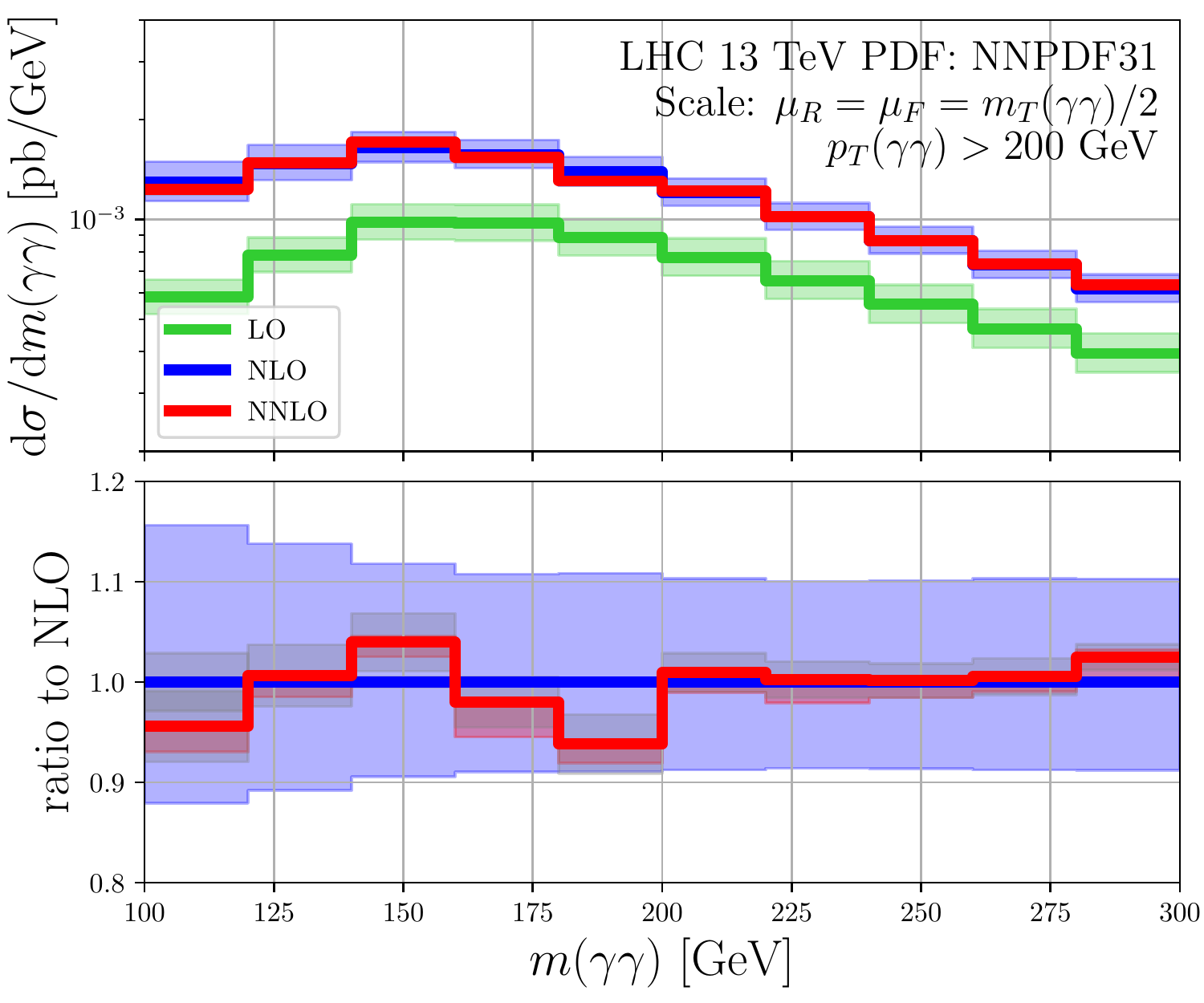}
\caption{As in fig.~\ref{fig:PT-m} but for the $m(\g\g)$ distribution subjected to different $\pt2g$ cuts: $\pt2g>50$ GeV (left), $\pt2g>100$ GeV (center) and $\pt2g>200$ GeV (right).}
\label{fig:m-cuts}
\end{centering}
\end{figure}

In this work we calculate the NNLO QCD corrections to a number of one-dimensional distributions in the following variables: the transverse momentum of the photon pair $p_T(\g\g)$, the invariant mass of the two photons $m(\g\g)$, the angle between the two photons in the Collins-Soper frame $\phi_{CS}$, the absolute difference in rapidities of the two photons $\Delta y(\g\g) = |y(\gamma_1) - y(\gamma_2)|$, the azimuthal angle between the two photons $\Delta\phi(\g\g)$ and the absolute rapidity of the photon pair $|y(\gamma\gamma)|$. We also calculate the NNLO QCD corrections to the following two-dimensional distributions: $m(\g\g)\otimes p_T(\g\g)$ and $\phi_{CS} \otimes m(\g\g)$.

We first discuss the $\pt2g$ differential distribution which is of central interest to this work. The distribution is shown in fig.~\ref{fig:PT-m}. As can be seen from this figure, the NLO QCD correction is very significant relative to the LO one. In particular, the scale uncertainty bands at LO and NLO do not overlap anywhere. This behavior is easy to understand based on the properties of inclusive diphoton production through NNLO. Clearly, a reliable prediction of this observable requires the inclusion of, at least, the NNLO QCD corrections. 

As can be seen from fig.~\ref{fig:PT-m} the inclusion of the NNLO corrections has a major stabilizing impact on the $\pt2g$ distribution. With the exception of the very low $\pt2g$ region which we will discuss shortly, the scale uncertainty decreases significantly, by a factor of about four, relative to NLO. Moreover the NLO and NNLO scale bands now overlap everywhere. Such a behavior is consistent with the expected stabilization of the inclusive diphoton production cross-section starting at N$^3$LO. 

The low-$\pt2g$ behavior of this distribution deserves special attention. A fixed-order perturbative description would not be adequate for $\pt2g$ below about 20 GeV due to the importance of resummation effects, but we expect it to be reliable for larger $\pt2g$ values. For this reason one may wonder why the low-$\pt2g$ part of the spectrum shows significantly increased scale dependence and larger NNLO/NLO $K$-factor. This behavior may be influenced by resummation effects however we do not expect them to be the dominant ones. We suspect that the main factor behind it is the loop-induced contribution $gg\to g\2g$ which only starts to contribute at NNLO. 

\begin{figure}[t]
\begin{centering}
\includegraphics[width = 0.8\textwidth,trim=0 1mm 0 0]{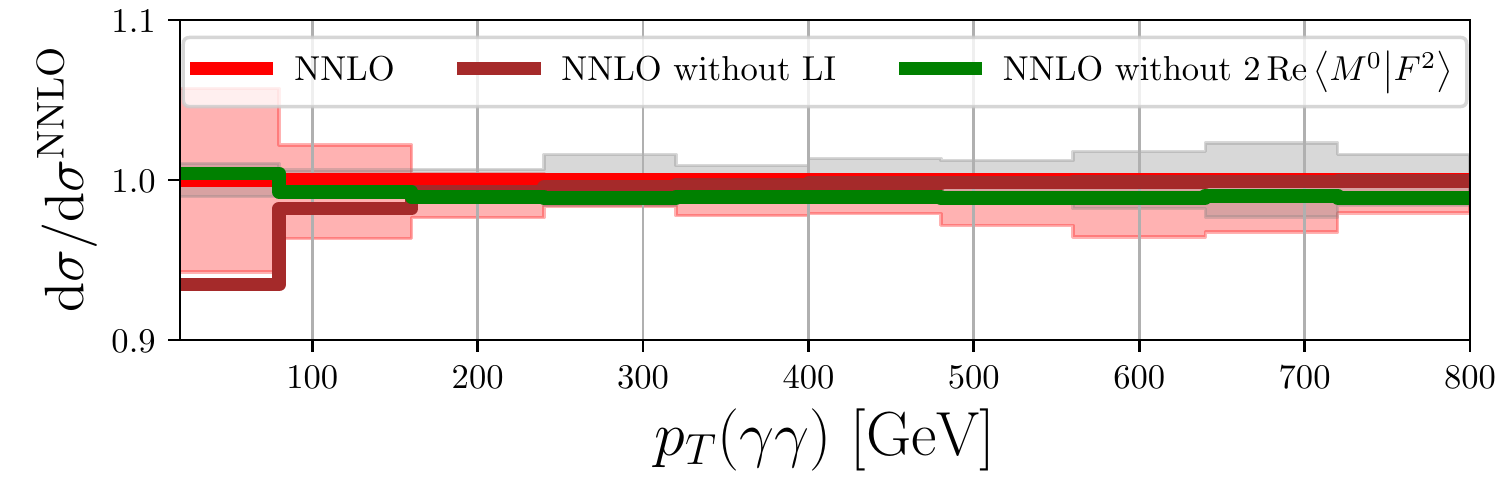}\\
\includegraphics[width = 0.48\textwidth,trim=0 1mm 0 0]{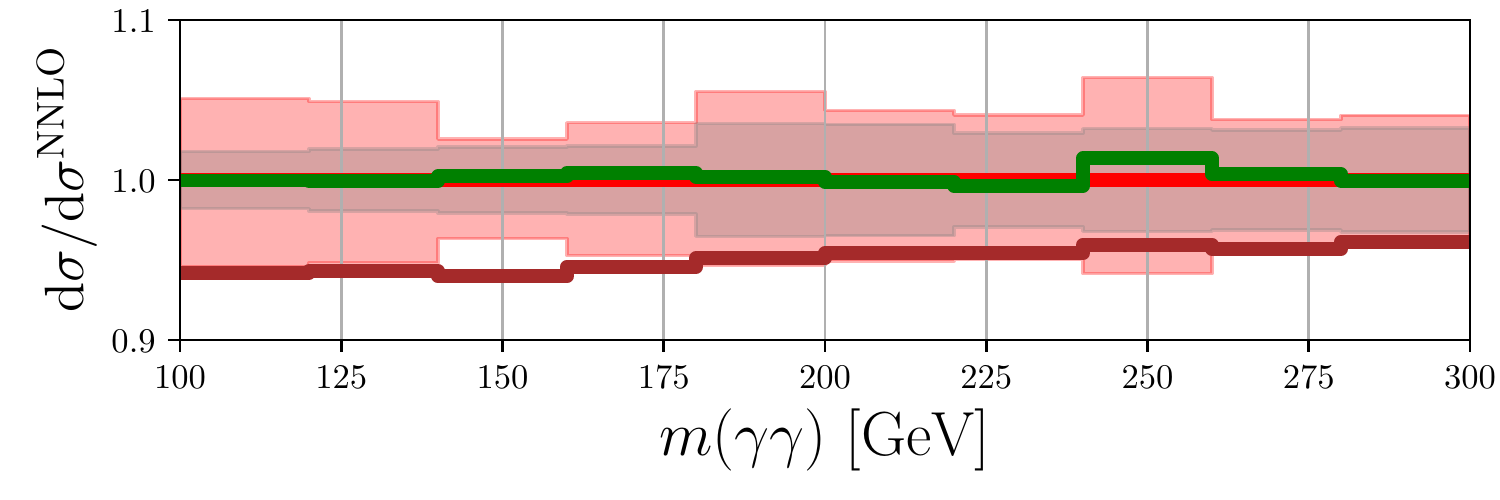}
\includegraphics[width = 0.48\textwidth,trim=0 1mm 0 0]{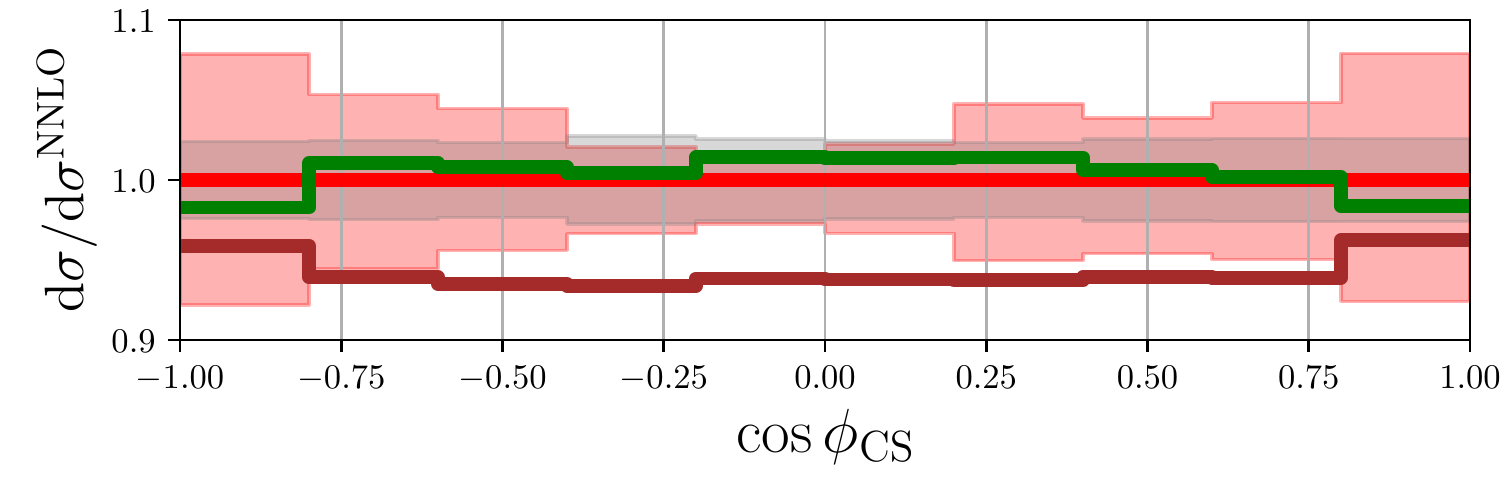}\\
\includegraphics[width = 0.48\textwidth,trim=0 1mm 0 0]{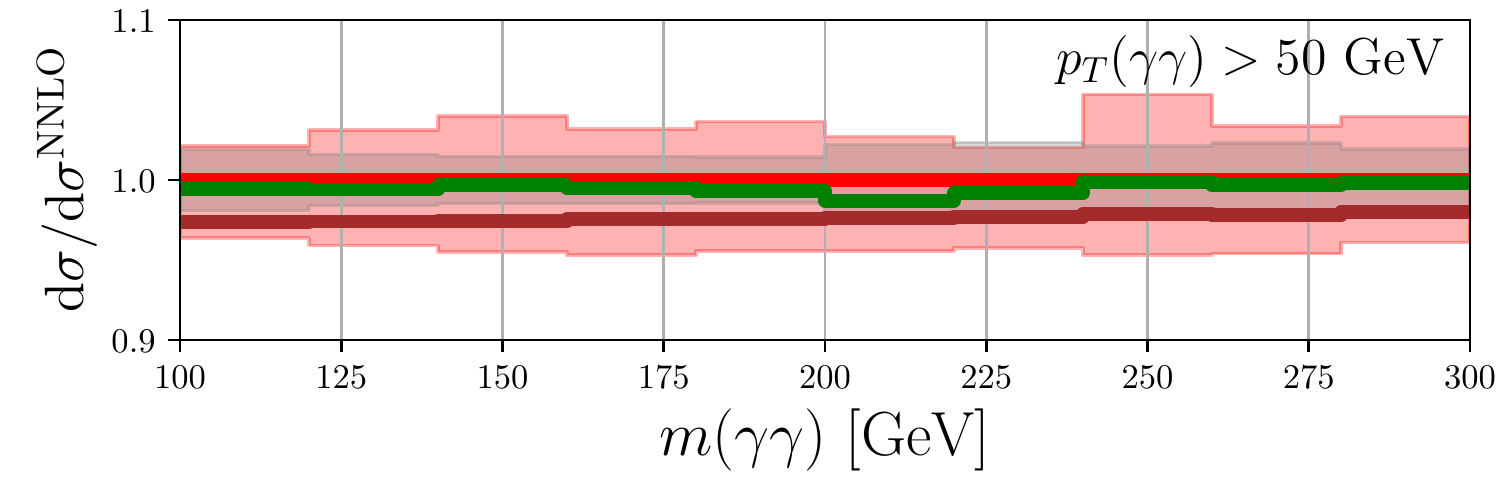}
\includegraphics[width = 0.48\textwidth,trim=0 1mm 0 0]{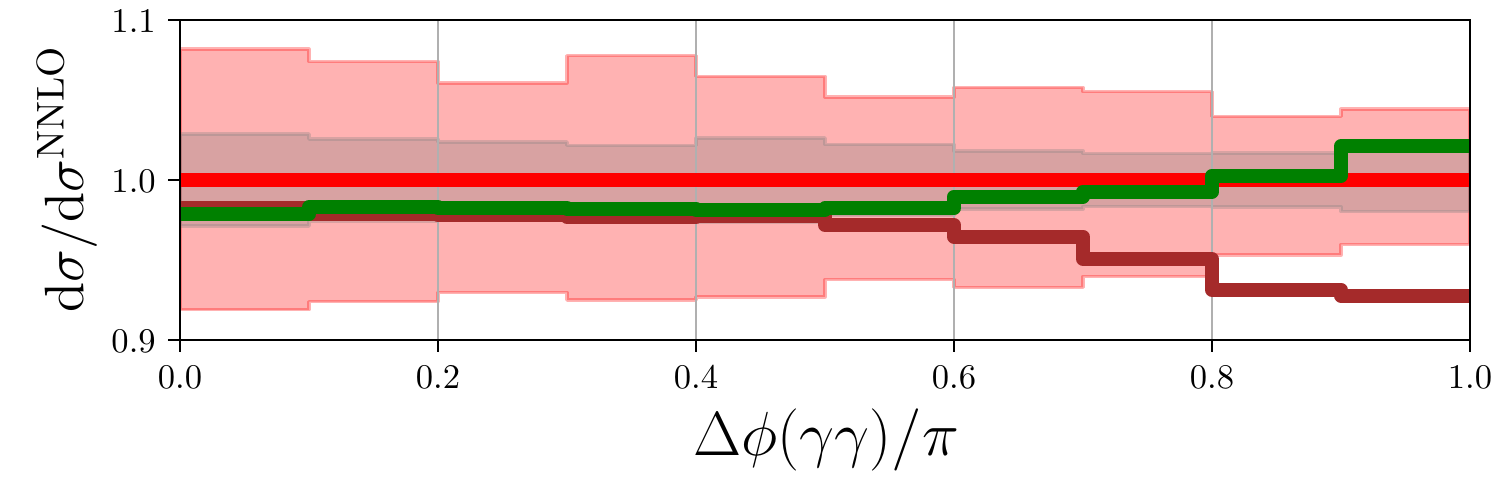}\\
\includegraphics[width = 0.48\textwidth,trim=0 1mm 0 0]{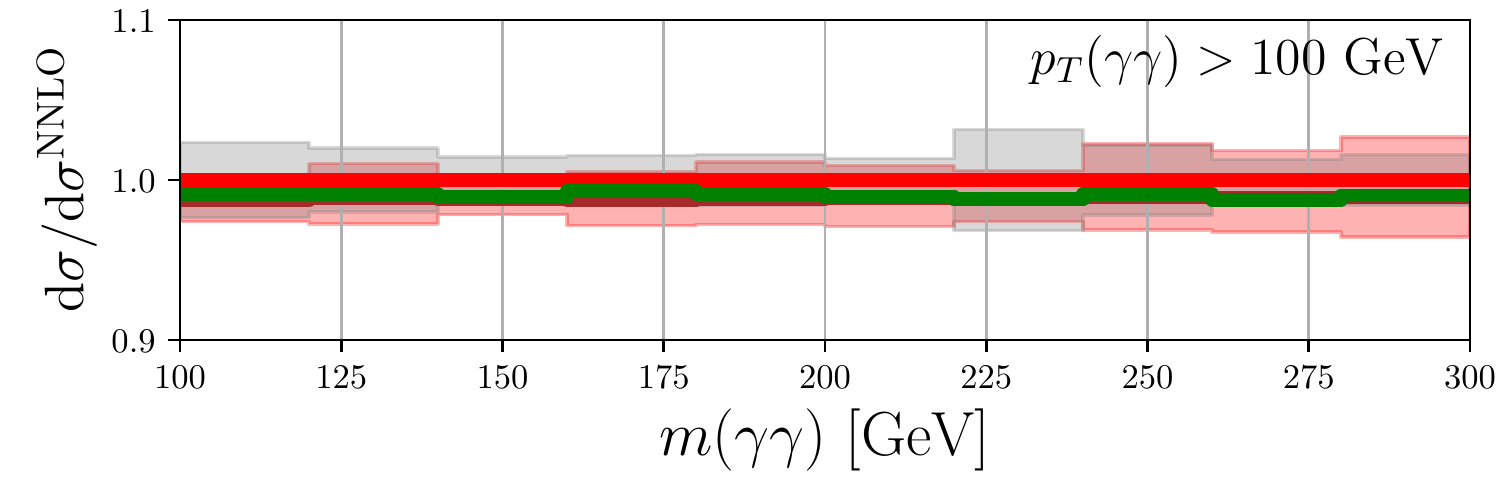}
\includegraphics[width = 0.48\textwidth,trim=0 1mm 0 0]{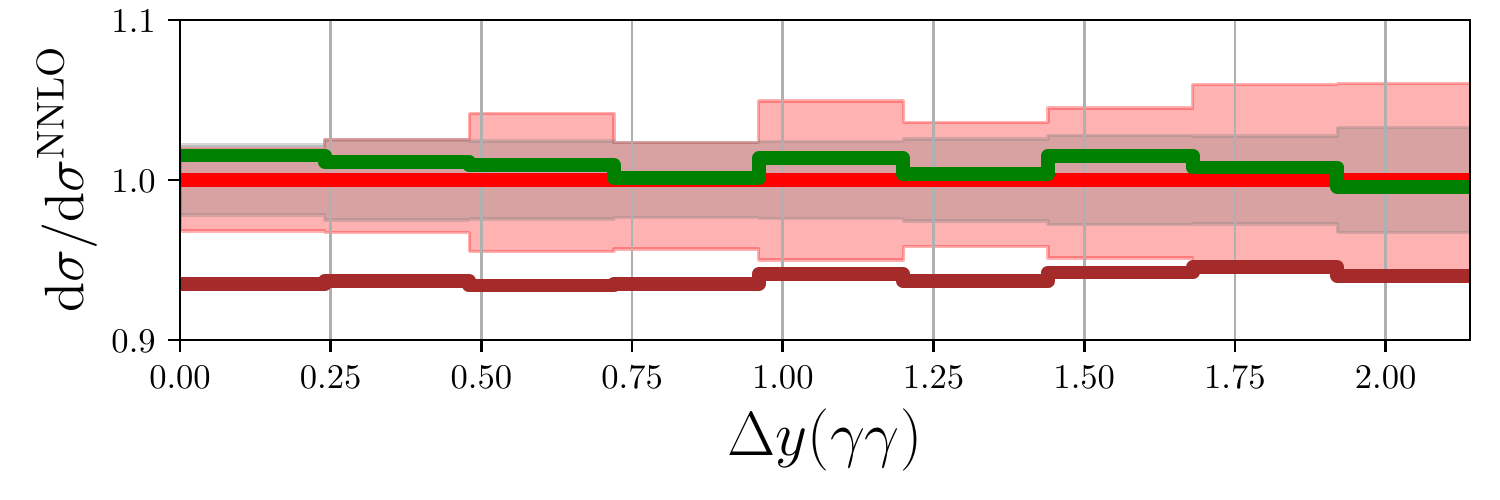}\\
\includegraphics[width = 0.48\textwidth,trim=0 1mm 0 0]{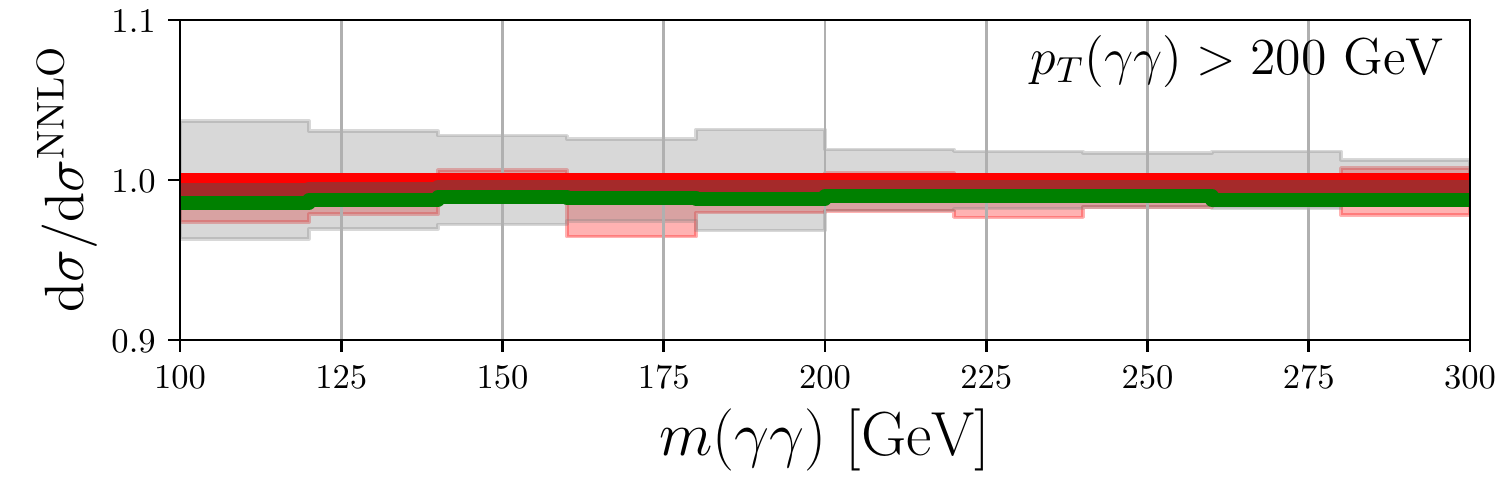}
\includegraphics[width = 0.48\textwidth,trim=0 1mm 0 0]{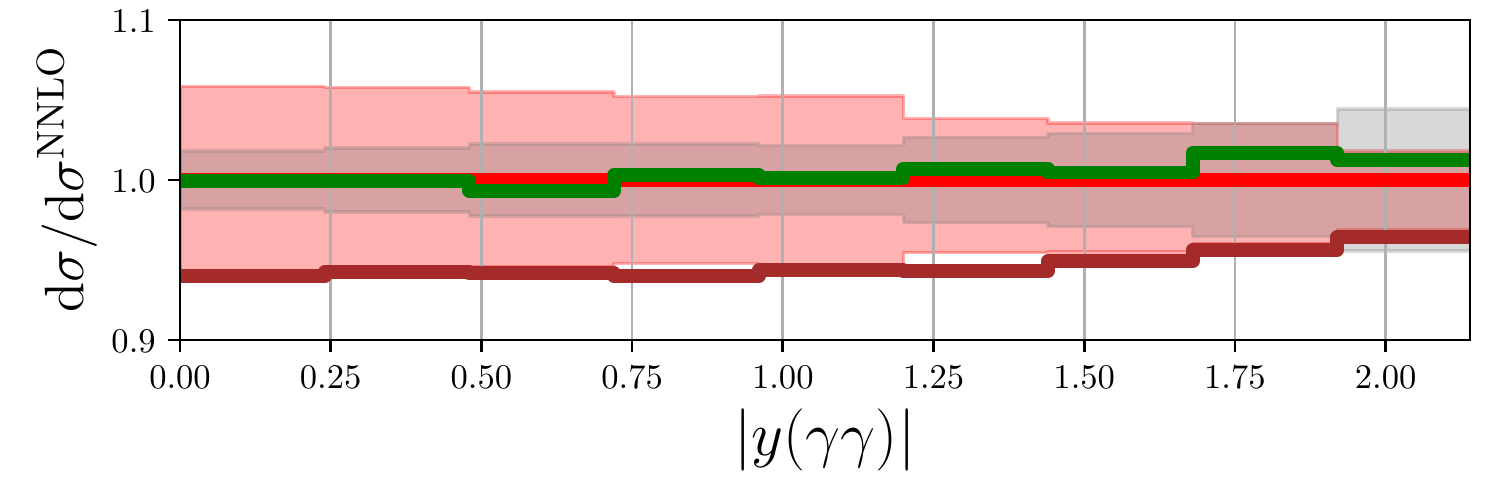}\\
\caption{Comparison of various approximations to the NNLO differential predictions: the complete NNLO prediction (red), NNLO excluding the loop-induced contribution (dark brown) and NNLO excluding the finite remainder ${\cal H}^{(2)}(s_{12})$ defined in eq.~(\ref{eq:amp-scale}) (green). The grey band shows the MC integration error of the complete NNLO prediction while the red band shows its scale variation.}
\label{fig:ratios}
\end{centering}
\end{figure}

The impact of this contribution is shown in fig.~\ref{fig:ratios} as a ratio of the full NNLO to the NNLO excluding this contribution. As fig.~\ref{fig:ratios} indicates the loop-induced correction is concentrated at relatively low $\pt2g$ values and becomes completely negligible for $\pt2g$ values about 200 GeV or larger. If the loop induced correction is excluded, the scale dependence of the first bin becomes smaller by about a factor of two and the NNLO/NLO $K$-factor also decreases by a factor of about two. 

Our findings indicate that at the level of NNLO QCD corrections, the loop-induced contribution becomes significant. This contribution can be tamed further, by including the NLO QCD correction to the loop-induced contribution (which is a partial N$^3$LO contribution for this process). Such a calculation requires the two-loop amplitude for the process $gg\to g\2g$. This result is not available in the literature but it is certainly within reach given the number of other five-point two-loop QCD amplitudes that have been computed.

Overall, the scale uncertainty of the $\pt2g$ spectrum at NNLO is rather small - about couple of percent for diphoton $p_T$'s above 200 GeV or so. This implies that this observable is well described within perturbative QCD and can be used in precision analyses of Higgs physics and searches for resonances decaying to diphoton final states. The dominant uncertainty at large $\pt2g$ is due to the Monte Carlo integration error of the calculation itself. It can be further improved albeit at a significant computational cost. A future update may also include pdf uncertainties, electroweak effects as well as the NLO correction to the loop-induced process. Finally, the effects from the the leading color approximation used here may also need to be improved upon. As can be seen from fig.~\ref{fig:ratios} the approximated contribution is a rather small 1-2\% effect and is much smaller than the MC error. It can also be improved upon once the complete two-loop amplitudes for this process become available.

\begin{figure}[t]
\begin{centering}
\includegraphics[width = 0.31\textwidth,trim=0 1mm 0 0]{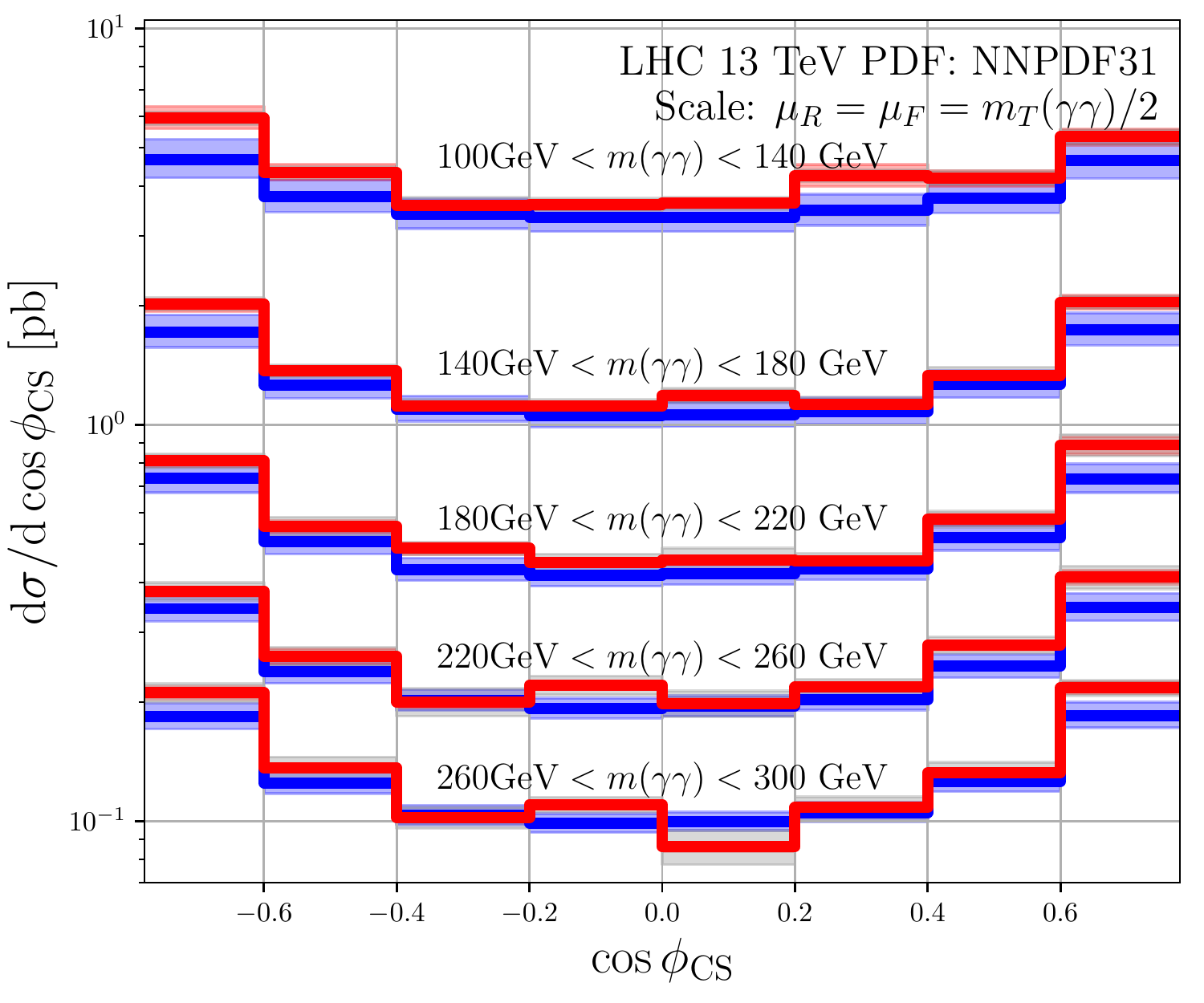}
\includegraphics[width = 0.31\textwidth,trim=0 1mm 0 0]{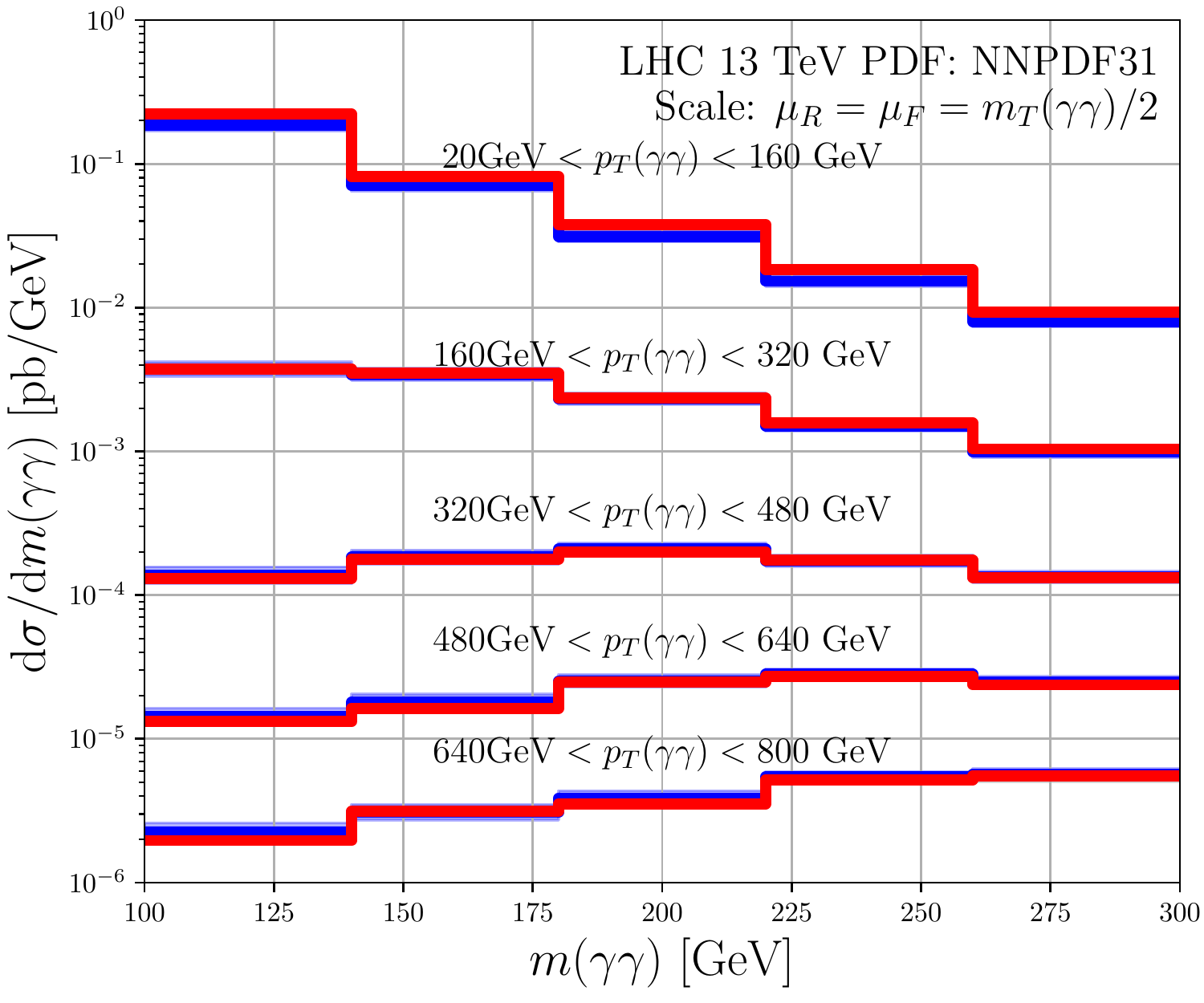}
\includegraphics[width = 0.31\textwidth,trim=0 1mm 0 0]{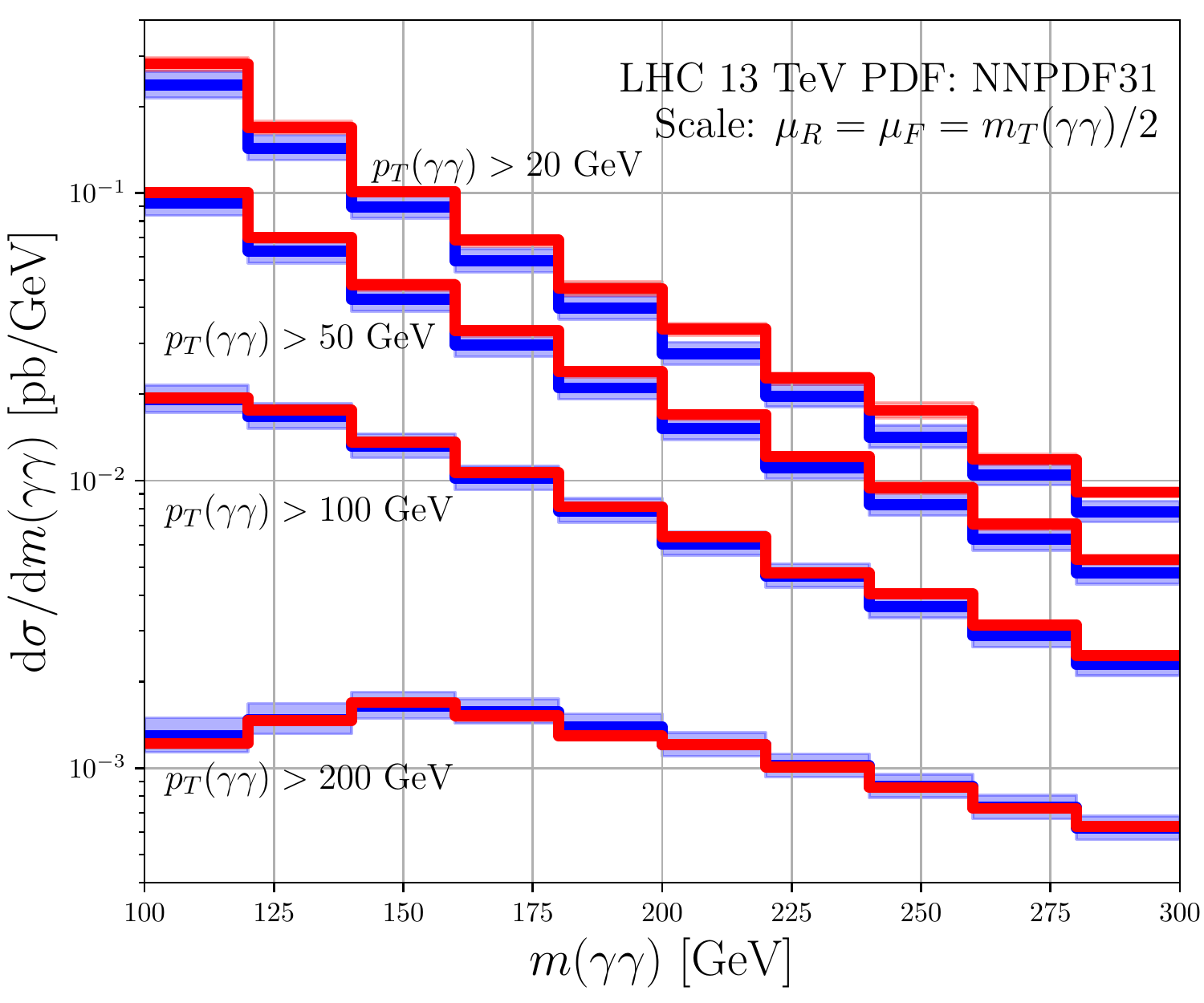}
\caption{Two-dimensional differential distributions: in $\phi_{CS} \otimes m(\g\g)$ (left) and in $m(\g\g)\otimes p_T(\g\g)$ but shown in two alternative forms and for different choice of bins: in $\pt2g$ bins (center) and with $\pt2g$ cuts (right). Only the NLO and NNLO central scale predictions and scale variation bands are shown. Note that the figure to the right shows the same results that already appear in figs.~\ref{fig:PT-m},\ref{fig:m-cuts}. }
\label{fig:2D}
\end{centering}
\end{figure}

We next turn our attention to the $\mgg$ distribution. It is shown in figs.~\ref{fig:PT-m} and \ref{fig:m-cuts}. This distribution is significant for any search of resonances decaying to diphotons at non-zero $p_T$. To thoroughly understand the interplay between $\mgg$ and $\pt2g$ we have shown the $\mgg$ distribution in several ways. In fig.~\ref{fig:PT-m} we show the $\mgg$ distribution subjected only to our default selection cuts. In fig.~\ref{fig:m-cuts} we show the same distribution but with more stringent $\pt2g$ cuts of 50 GeV, 100 GeV and 200 GeV. A summary of the same result (only the NLO and NNLO are displayed) is shown in fig.~\ref{fig:2D} (right). Fig.~\ref{fig:2D} (center) shows the $\mgg$ distributions for several slices of $\pt2g$. 

From these plots it is clear that the $\mgg$ distribution has a pattern of higher-order corrections that is similar to $\pt2g$: large NLO/LO corrections and much smaller NNLO/NLO ones. The size of the NNLO corrections strongly depends on the $\pt2g$ cut and they decrease as the cut increases. For small values of the $\pt2g$ cut the NNLO and NLO scale bands do not overlap while they fully overlap for $\pt2g$ cuts above 100 GeV. One may wonder if such a behavior is related to the loop-induced contribution. In fig.~\ref{fig:ratios} we have shown its effect for all $\pt2g$ cuts considered in this work. From this one can conclude that indeed the size of the loop-induced correction is consistent with the non-overlap of the NLO and NNLO scale bands. This means that for theoretical predictions to be reliable with full NNLO accuracy for $\pt2g$ cuts below 100 GeV or so, the NLO corrections to the loop induced contributions might need to be included. 

In general, the effect of the NNLO correction on the $\mgg$ distribution is a rather flat shift with respect to the NLO one and leads to a decrease of the scale uncertainty by a factor of about two at low $\pt2g$ and four or more at large $\pt2g$. Another important source of error is the MC integration one. The effect from the leading color approximation in the finite remainder is at the percent level and therefore insignificant. 

\begin{figure}[t]
\begin{centering}
\includegraphics[width = 0.48\textwidth,trim=0 1mm 0 0]{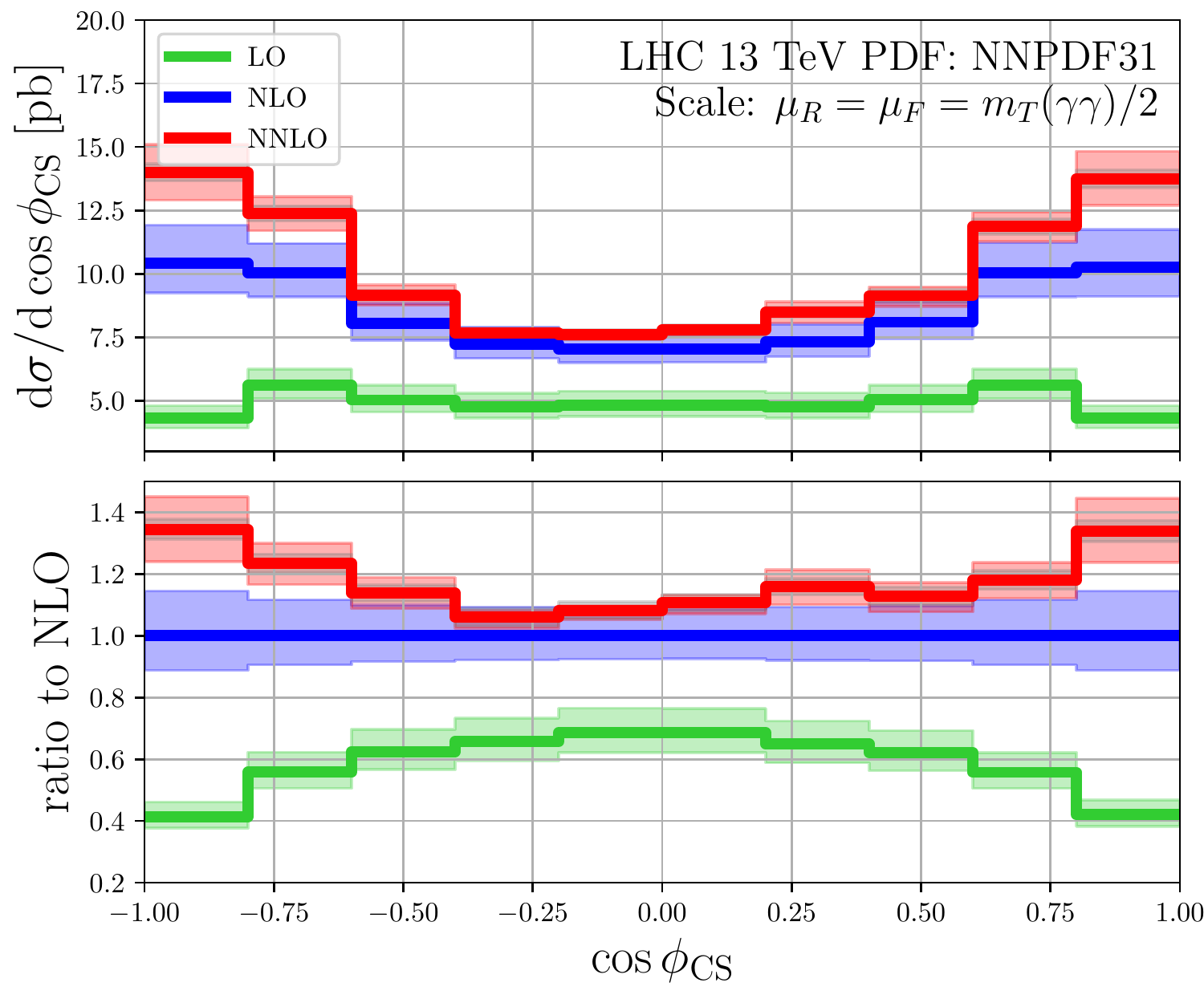}
\includegraphics[width = 0.48\textwidth,trim=0 1mm 0 0]{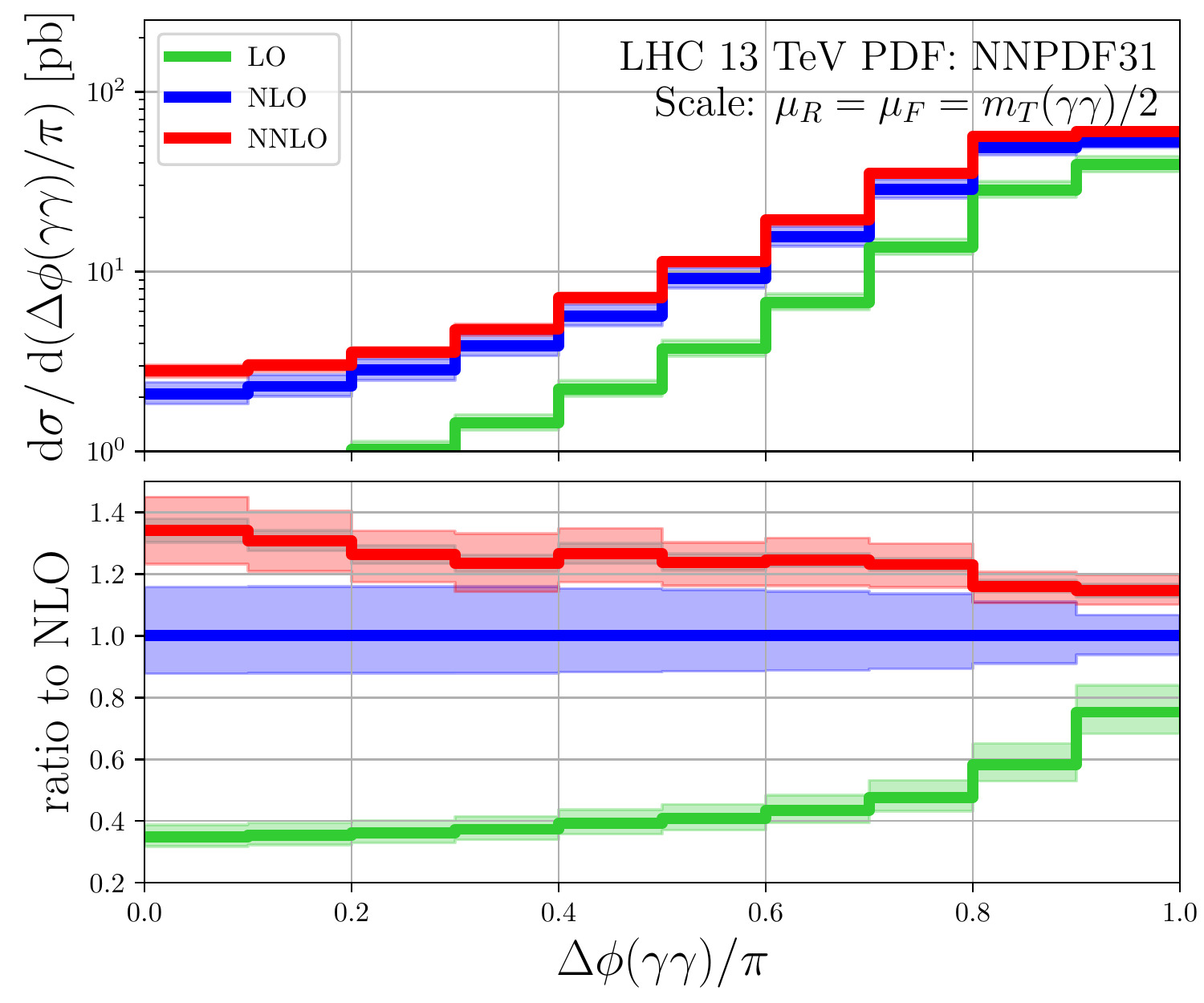}
\caption{As in fig.~\ref{fig:PT-m} but for the angular distributions in $\phi_{CS}$ (left) and $\Delta\phi(\g\g)$ (right).}
\label{fig:angles}
\end{centering}
\end{figure}

\begin{figure}[t]
\begin{centering}
\includegraphics[width = 0.48\textwidth,trim=0 1mm 0 0]{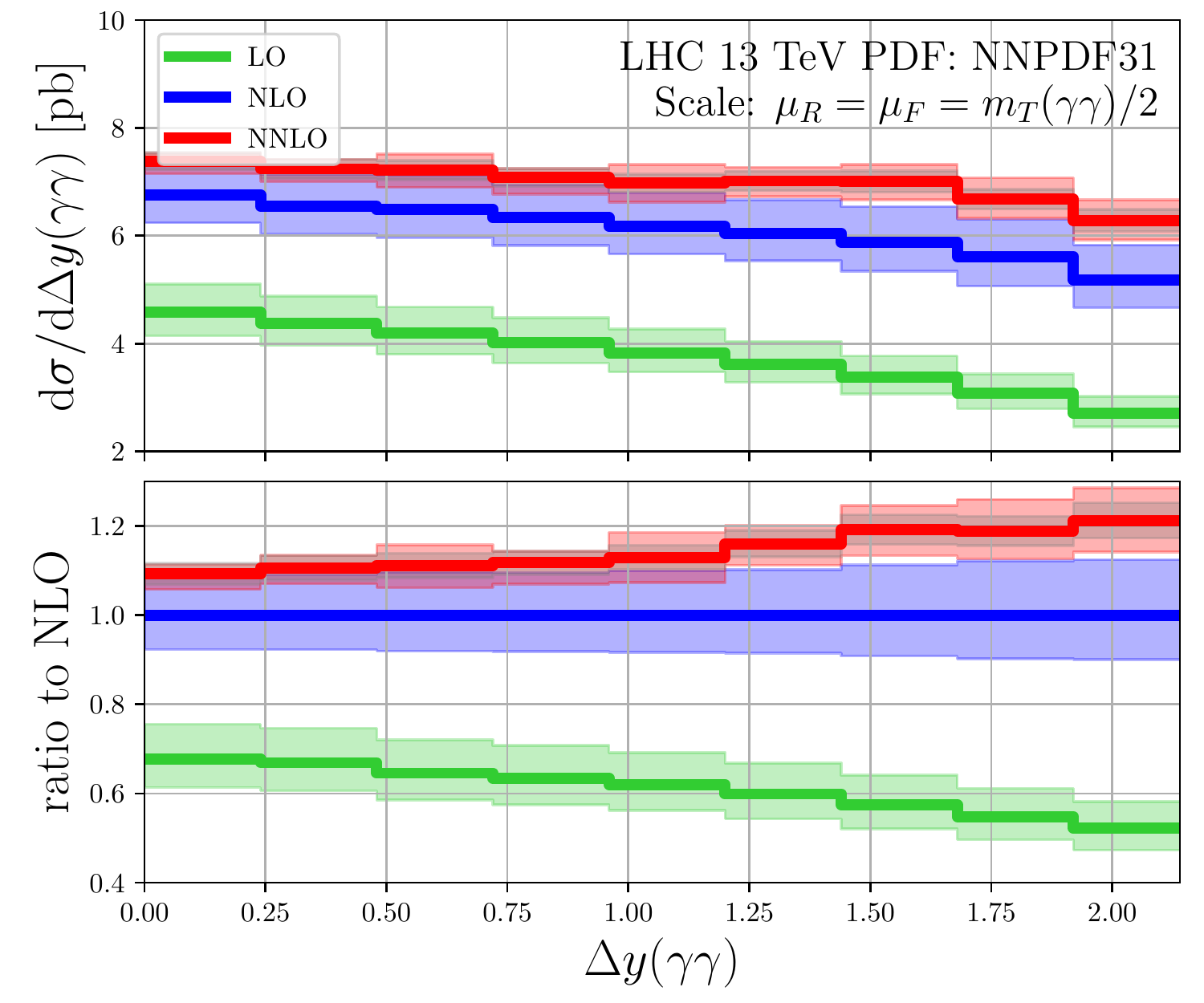}
\includegraphics[width = 0.48\textwidth,trim=0 1mm 0 0]{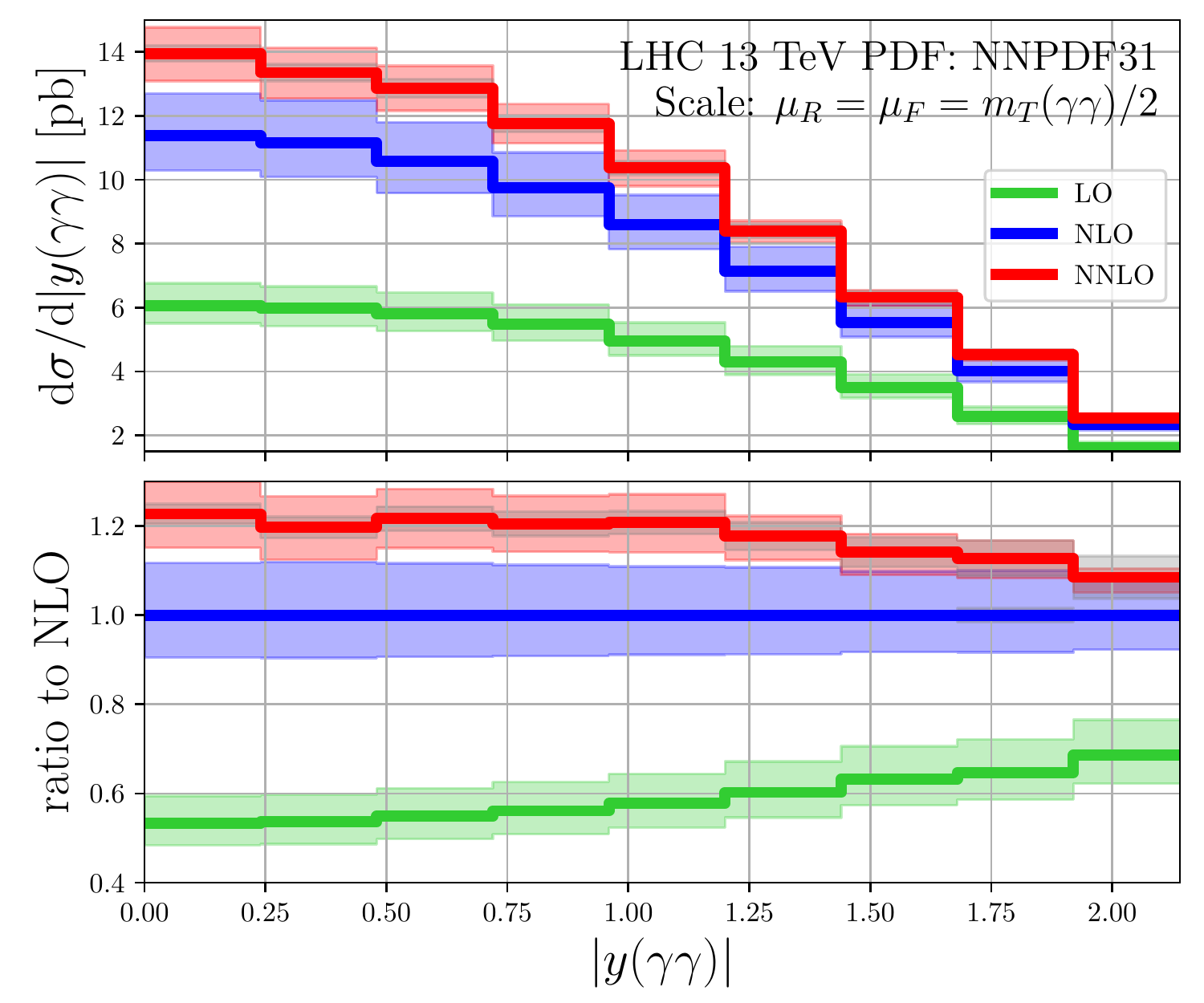}
\caption{As in fig.~\ref{fig:PT-m} but for the following rapidity distributions: $\Delta y(\g\g)$ (left) and $|y(\gamma\gamma)|$ (right).}
\label{fig:y}
\end{centering}
\end{figure}

In fig.~\ref{fig:angles} we show distributions in the angular variables $\phi_{CS}$ and $\Delta\phi(\g\g)$ while in fig.~\ref{fig:y} we show the $\Delta y(\g\g)$ and $|y(\gamma\gamma)|$ rapidity distributions. The $\phi_{CS}$ distribution in slices of $\mgg$ is shown in fig.~\ref{fig:2D} (left). All these distributions have very large NLO/LO $K$-factors. Unlike the $\pt2g$ and $\mgg$ distributions, however, they also have sizable NNLO corrections which in most bins are outside the NLO uncertainty bands. This pattern of higher order corrections indicates that for the scale used in this work, the NLO approximation is inadequate for describing these distributions. 

\begin{figure}[t]
\begin{centering}
\includegraphics[width = 0.8\textwidth,trim=0 1mm 0 0]{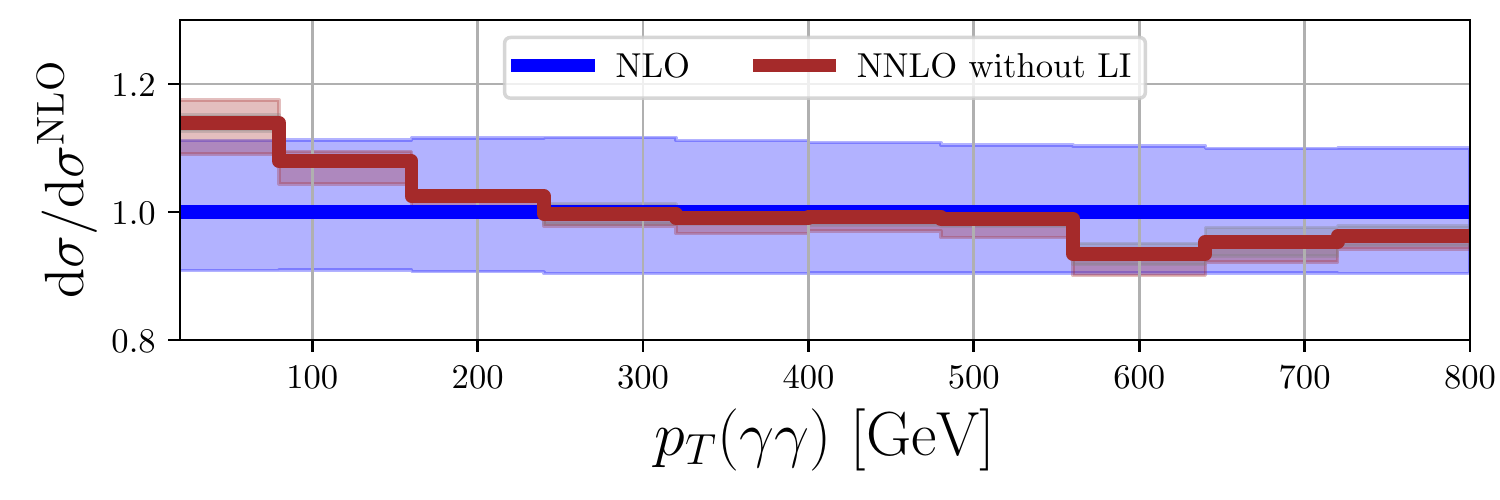}\\
\includegraphics[width = 0.48\textwidth,trim=0 1mm 0 0]{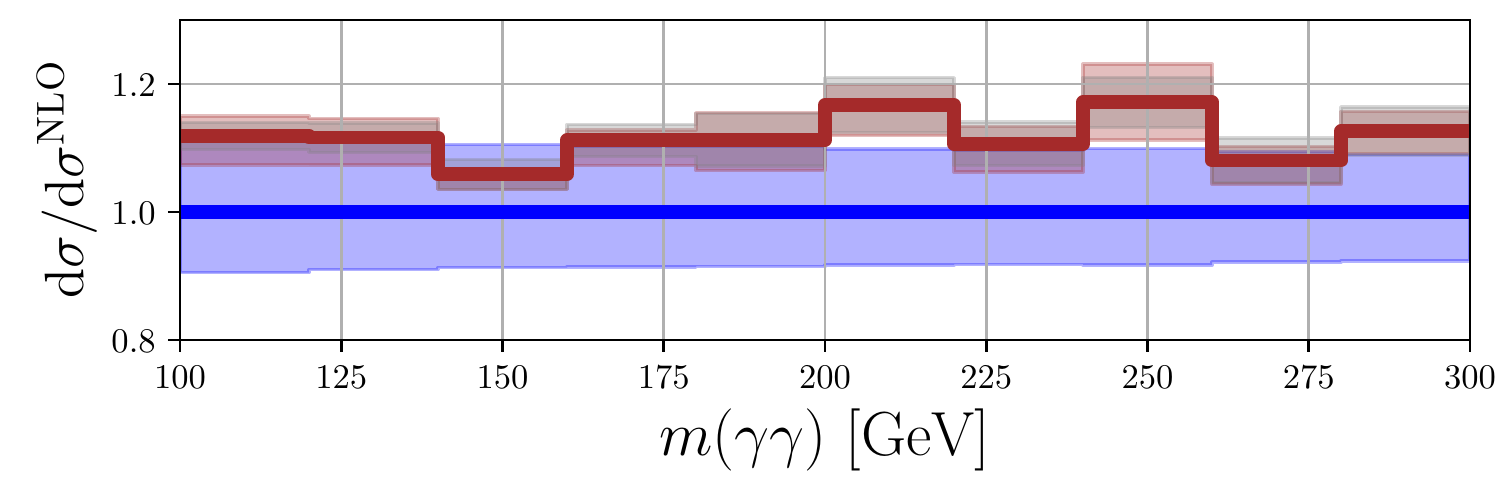}
\includegraphics[width = 0.48\textwidth,trim=0 1mm 0 0]{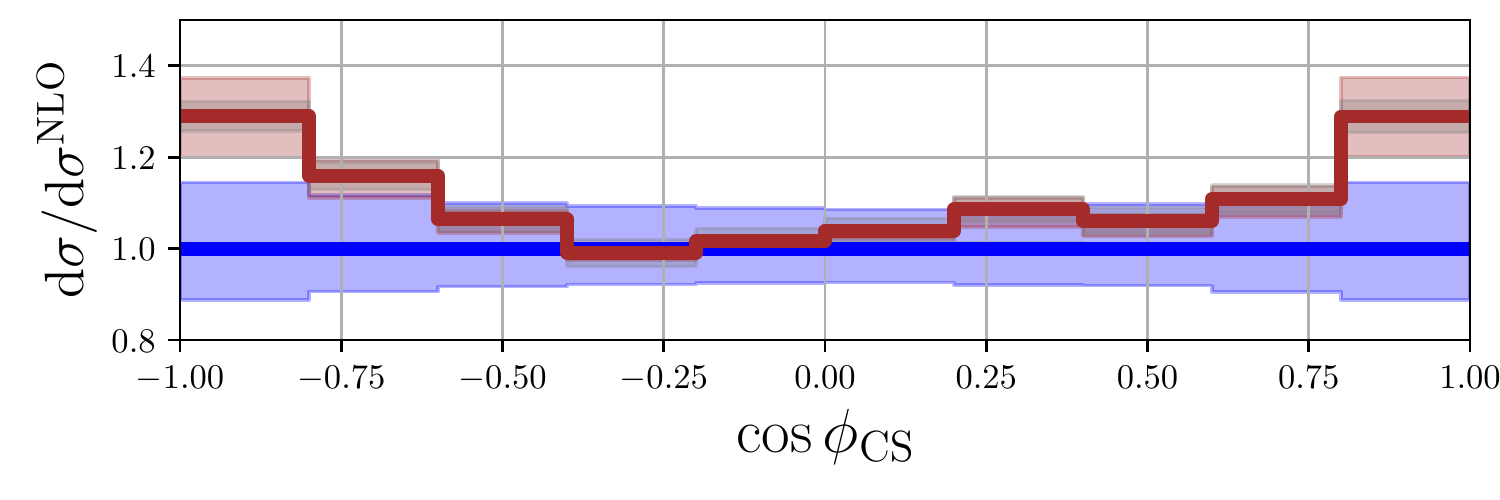}\\
\includegraphics[width = 0.48\textwidth,trim=0 1mm 0 0]{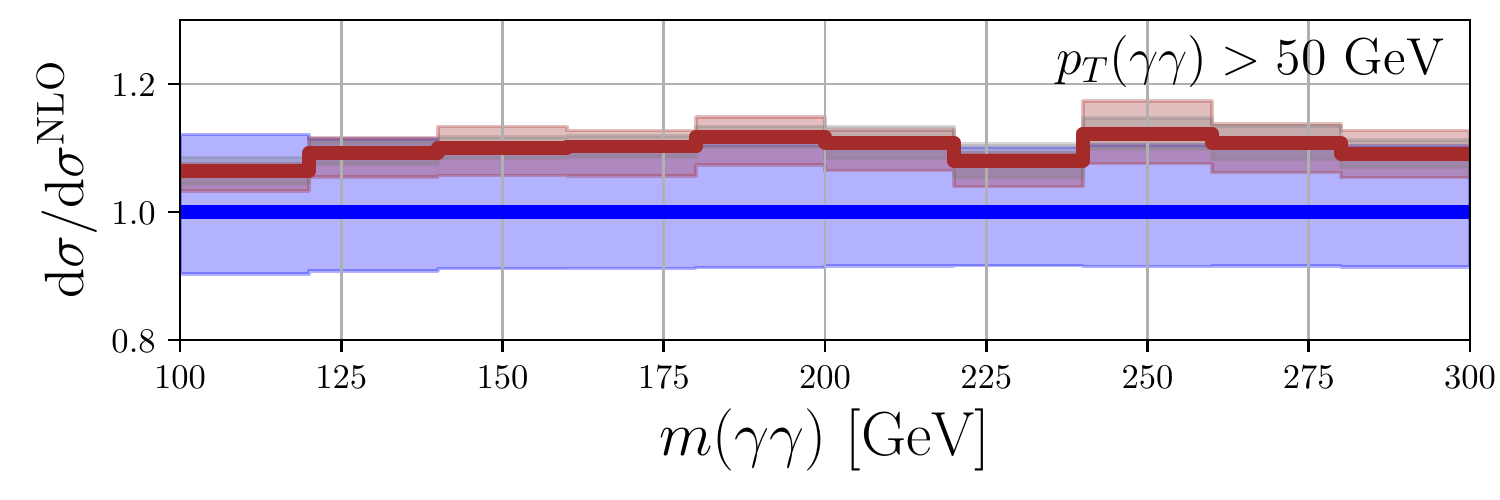}
\includegraphics[width = 0.48\textwidth,trim=0 1mm 0 0]{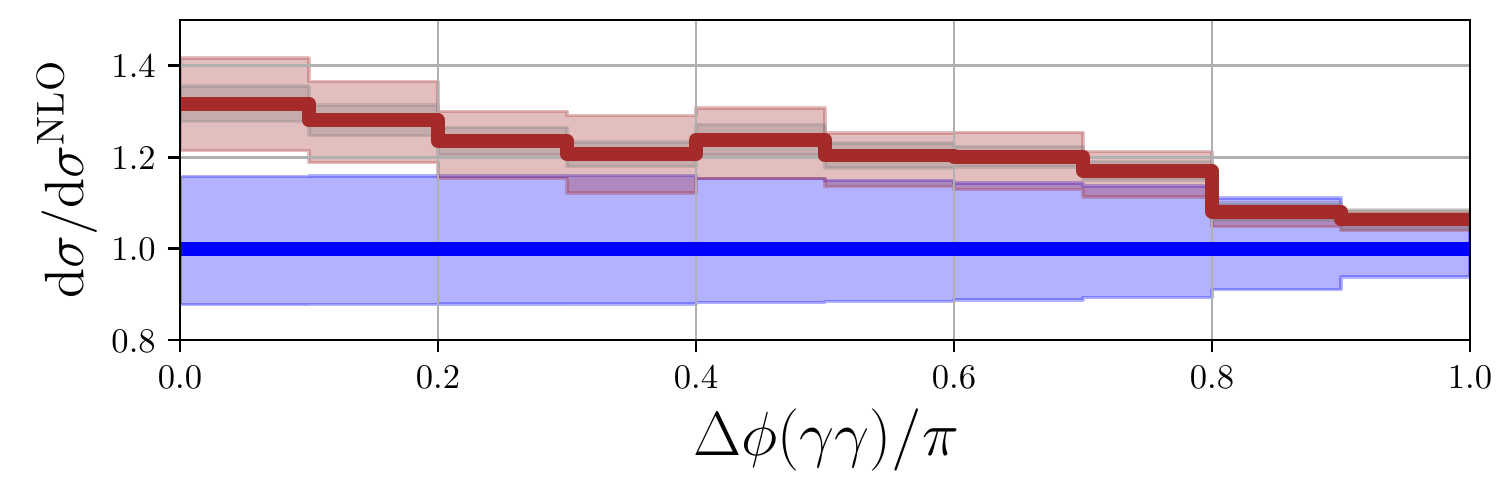}\\
\includegraphics[width = 0.48\textwidth,trim=0 1mm 0 0]{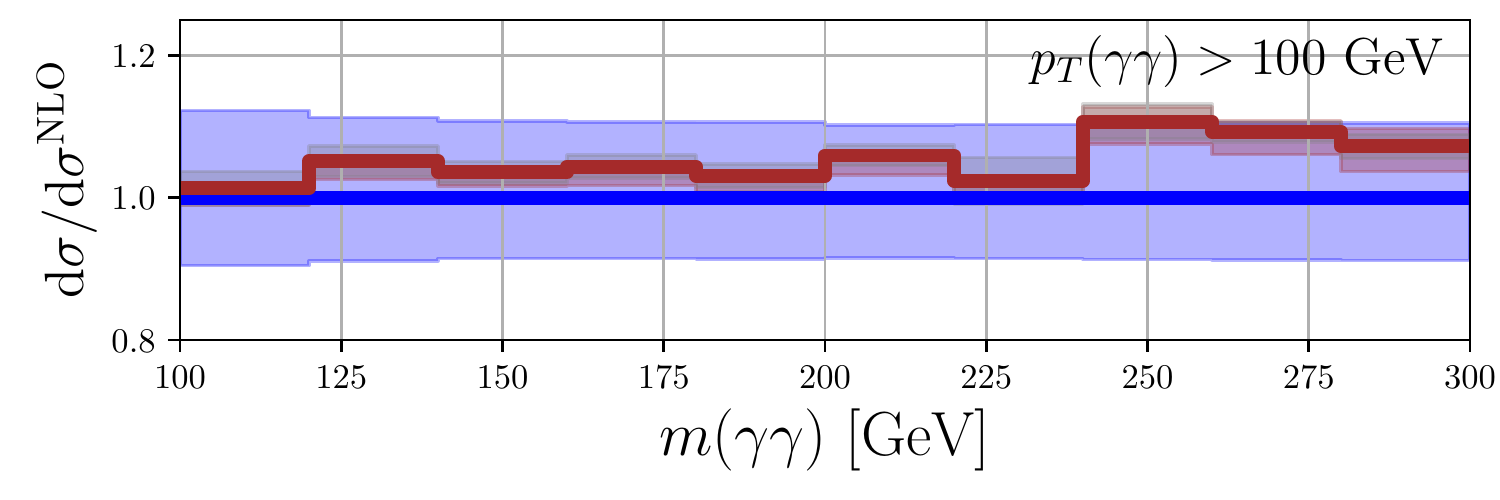}
\includegraphics[width = 0.48\textwidth,trim=0 1mm 0 0]{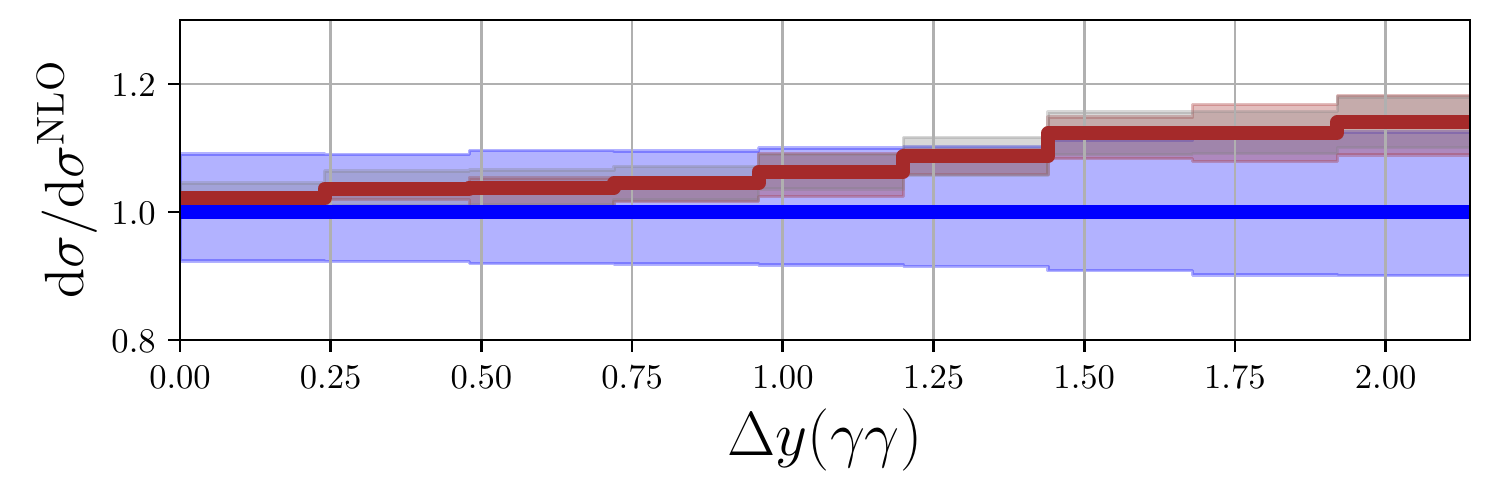}\\
\includegraphics[width = 0.48\textwidth,trim=0 1mm 0 0]{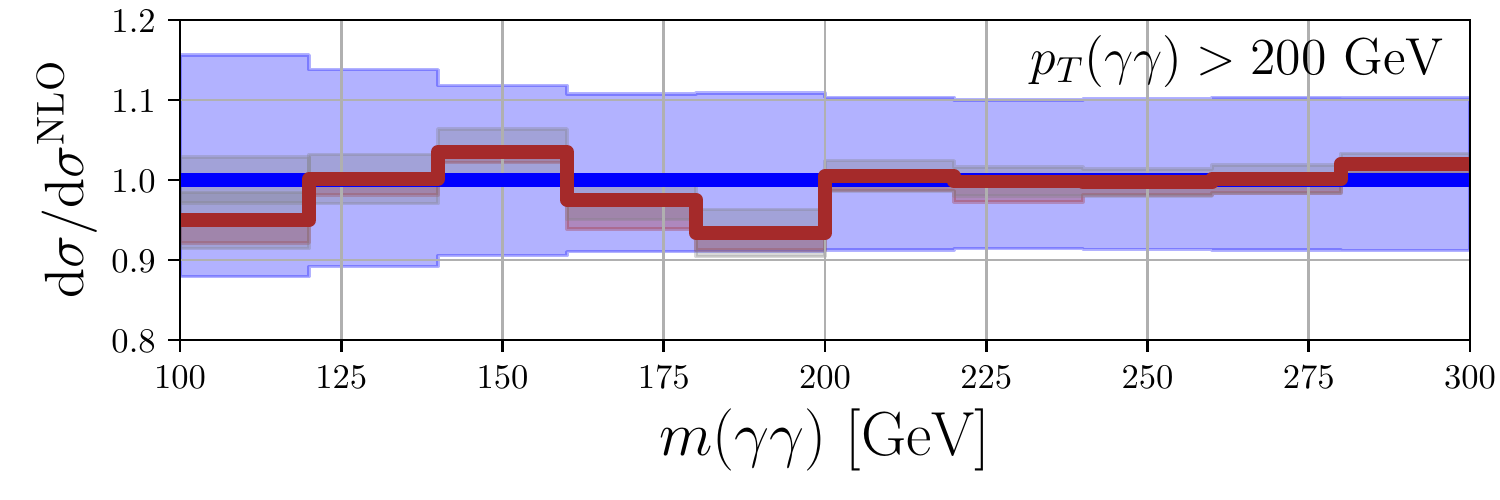}
\includegraphics[width = 0.48\textwidth,trim=0 1mm 0 0]{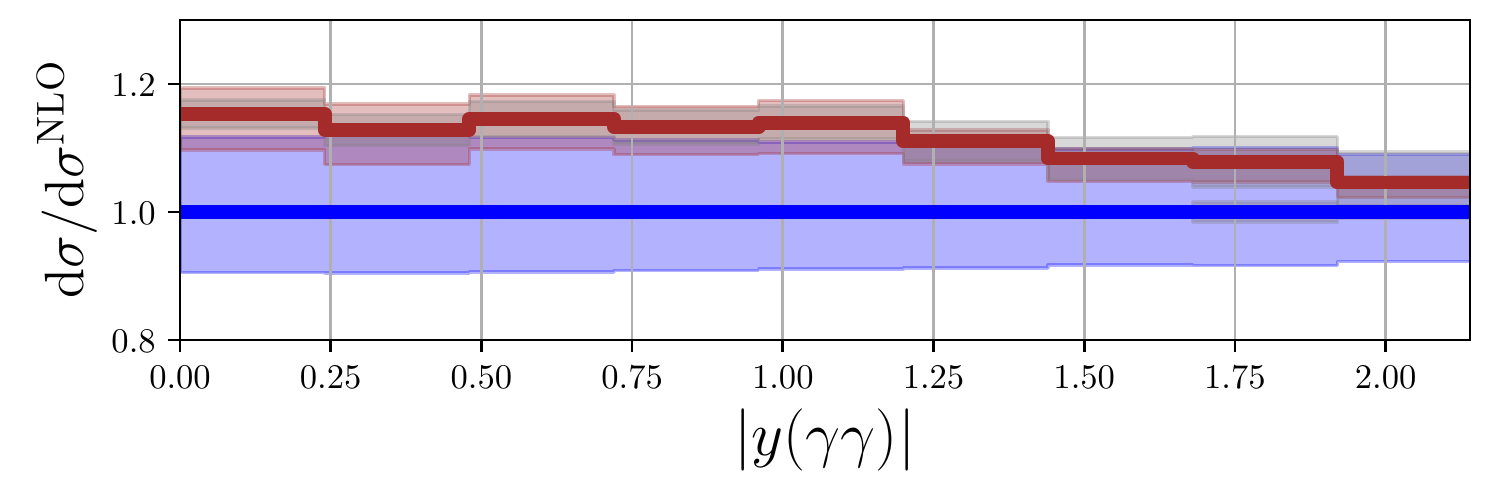}\\
\caption{Comparison of NLO (blue) scale bands to NNLO QCD excluding the loop-induced contribution (dark brown).}
\label{fig:ratios2}
\end{centering}
\end{figure}

Based on the above observations one may question the presence of perturbative stability in these variables. As a first step towards analyzing this we consider the behavior of the NNLO prediction without the loop-induced contribution (in the following we refer to it as NNLO-minus-LI). The numerical impact of the loop-induced contribution for each differential distribution can be seen in fig.~\ref{fig:ratios} and fig.~\ref{fig:ratios2}. We observe the following. For the $\phi_{CS}$ distribution the NNLO-minus-LI scale uncertainty band is mostly within the NLO one or the two bands overlap. This is not the case for the first and last bins of this distribution, however, the behavior of the $\phi_{CS}$ distribution in these two bins is strongly affected by the kinematic cuts. The NNLO-minus-LI band for the $\Delta y(\g\g)$ distribution overlaps in all bins with the NLO one. Same can be observed for the case of the $|y(\gamma\gamma)|$ distribution. In fact, the only distribution for which the NNLO-minus-LI and NLO scale uncertainty bands do not mostly overlap is the $\Delta\phi(\g\g)$ one. For this distribution we observe that the NNLO-minus-LI and NLO scale uncertainty bands overlap for $\Delta\phi(\g\g)/\pi>0.6$ while below this value they are not very far apart, see fig.~\ref{fig:ratios2}. Given that the NLO/LO $K$-factor in this region is more than a factor of two it seems that such a non-overlap is not too concerning. 

From the above discussion it seems reasonable to conclude that the non-overlap between NNLO and NLO scale uncertainty bands observed in the angular and rapidity diphoton distributions is somewhat affected by the loop-induced contribution. It is therefore plausible to assume that the inclusion of this contribution's NLO correction may alleviate this non-overlap. Other factors that may be affecting this behavior is the choice of scale as well as resummation effects which are relevant at low $\pt2g$. A detailed investigation of those is however outside the scope of this work. On the other hand, as can also be seen from fig.~\ref{fig:ratios}, the two-loop finite remainder has a rather small contribution and, therefore, we do not expect these distributions to be significantly affected by two-loop subleading color corrections.


\section{Conclusions}\label{sec:conclusions}

In this work we calculate the NNLO QCD corrections to the process $pp\to \2g+{\rm jet}$. This process is the main background to high-$p_T$ Higgs boson production decaying to two photons. The main result of this work is the calculation of the diphoton $p_T$ spectrum with NNLO accuracy. The NNLO correction to this variable is important and it brings the uncertainty from un-calculated higher-order corrections down to a couple of percent at intermediate and large values of $\pt2g$. Overall, the quality of the theoretical prediction for this distribution is very high and it appears to be under good theoretical control. The same conclusion applies for double differential distributions in $\pt2g$ and $\mgg$. 

We have suggested various possible avenues for further improving the quality of the theoretical predictions in this process. They include the calculations of the partial N$^3$LO corrections due to loop induced processes which can be calculated with the help of NLO technology. The only missing ingredient for such a calculation is the two-loop amplitude for the process $gg\to g\2g$ whose calculation is within reach. A more extensive study of possible scale choices for this process might also be beneficial given the very high precision reached in the $\pt2g$ distribution. Merging our fixed-order calculations with resummed calculations will allow for a quality description of the $\pt2g$ spectrum from very high down to very low values of $\pt2g$.

We conclude by stressing that the quality of the theoretical description achieved for this process is high which makes it possible to use it in background estimates for Higgs boson studies and related searches as well as in dedicated measurements of diphoton production. 

{\bf Note Added:} After the completion of the current work, ref.~\cite{Agarwal:2021vdh} appeared. It provides the subleading-colour expressions for the two-loop amplitudes for this process. We will include them in a future update of this work.

\begin{acknowledgments}
The work of M.C. was supported by the Deutsche Forschungsgemeinschaft under grant 396021762 - TRR 257. The research of A.M. and R.P. has received funding from the European Research Council (ERC) under the European Union's Horizon 2020 Research and Innovation Programme (grant agreement no. 683211). A.M. was also supported by the UK STFC grants ST/L002760/1 and ST/K004883/1. The research of H.C. is supported by the ERC Starting Grant 804394 HipQCD. A.M. acknowledges the use of the DiRAC Cumulus HPC facility under Grant No. PPSP226.
\end{acknowledgments}

\end{document}